\documentstyle[12pt]{article}
\begin{document}
\begin{titlepage}
\begin{center}
\hfill YITP-SB-02/63 \\
\vskip 20mm

{\Huge Four dimensional "old minimal" ${\cal N}=2$ supersymmetrization of 
${\cal R}^4$}
\vskip 10mm
Filipe Moura
\vskip 4mm
{\em C. N. Yang Institute for Theoretical Physics \\
State University of New York \\
Stony Brook, NY 11794-3840, U.S.A}\\
{\tt fmoura@insti.physics.sunysb.edu}
\vskip 6mm
\end{center}
\vskip .2in

\begin{center} {\bf Abstract } \end{center}
\begin{quotation}
\noindent
We write in superspace the lagrangian containing the fourth power of the Weyl 
tensor in the "old minimal" $d=4$, ${\cal N}=2$ supergravity, without local 
SO(2) symmetry. Using gauge completion, we analyze the lagrangian in 
components. We find out that the auxiliary fields which belong to the Weyl and 
compensating vector multiplets have derivative terms and therefore cannot be 
eliminated on-shell. Only the auxiliary fields which belong to the 
compensating nonlinear multiplet do not get derivatives and could still be 
eliminated; we check that this is possible in the leading terms of the 
lagrangian. We compare this result to the similar one of "old minimal" 
${\cal N}=1$ supergravity and we comment on possible generalizations to other 
versions of ${\cal N}=1, 2$ supergravity.

\end{quotation}
\vfill
\flushleft{\today}
\end{titlepage}
\eject

\section{Introduction}
\indent

The supersymmetrization of the fourth power of the Riemann tensor has been 
an active topic of research. In four dimensions, a term like this would be the 
leading bosonic contribution to a possible three-loop supergravity counterterm 
\cite{dks77}. In type I supergravity in ten dimensions, an ${\cal R}^4$ term is
necessary to cancel gravitational anomalies \cite{gs84, gs85}. ${\cal R}^4$ 
terms also show up in the low energy field theory effective action of both type
II and heterotic string theories, as was shown in \cite{gw86, gs87,gz86}.

In previous papers \cite{moura02,moura01} we have worked out the ${\cal N}=1$
supersymmetrization of ${\cal R}^4$ in four dimensions. We have shown that, in
the "old minimal" formulation with this term, the auxiliary fields $M, N$
could still be eliminated, but $A_m$ could not (it got derivative couplings in
the lagrangian that led to a dynamical field equation).

The goal of this article is to extend this supersymmetrization to ${\cal N}=2$ 
supergravity, which also admits off-shell formulations, and compare the result 
to the ${\cal N}=1$ one.

We start by briefly reviewing how one can obtain the "old minimal" (without
local SO(2)) ${\cal N}=2$ Poincar\'e supergravity from the conformal theory by
coupling to compensating vector and nonlinear multiplets. We then write, in 
superspace, using the known chiral projector and chiral density, a lagrangian 
that contains the fourth power of the Weyl tensor.
We start expanding this action in components and we quickly conclude that some 
of these auxiliary fields get derivatives and cannot be eliminated. For the
other auxiliary fields, we make a more detailed analysis and we show that 
their derivative terms cancel and, therefore, it should still be possible to 
eliminate them. We analyze, for which multiplet (Weyl, compensating vector and 
compensating nonlinear) which auxiliary fields can and which cannot be 
eliminated. We proceed analogously in the ${\cal N}=1$ case, using the previous
known results. We compare the two cases and discuss what can be generalized
to other versions of both these theories.

In appendix \ref{appendix2} we give a survey of curved SU(2) superspace, namely
its field content and the solution of the Bianchi identities.

\section{${\cal N}=2$ supergravity in superspace}
\setcounter{equation}{0}
\subsection{${\cal N}=2$ conformal supergravity in superspace}
\indent

Conformal supergravity theories were found for ${\cal N}\leq4$ in four 
dimensions. These theories have a local internal U($\cal N$) symmetry 
which acts on the supersymmetry gene\-rators $Q_A^a$ and $S_A^a$, with 
$a=1, \cdots, {\cal N}$; they can be formulated off-shell in conventional 
extended superspace, with structure group SO(1,3)$\times$U($\cal N$), where 
their actions, written as chiral superspace integrals, are known at the full 
nonlinear level. The first discussion of the conformal properties of extended 
superspace was given in \cite{gates80}; other later references are the seminal 
paper \cite{howe82} and the very nice review \cite{muller89}. Here we summarize
the main results we need.

In superspace, the main objects are the supervielbein $E_\Pi^M$ and the 
superconnection $\Omega_{\Lambda N}^{\ \ \ \ P}$ (which can be decomposed in 
its Lorentz and U($\cal N$) parts), in terms of which we write the torsions 
$T_{MN}^{\ \ \ P}$ and curvatures $R_{MN}^{\ \ \ PQ}$. Their arbitrary 
variations are given by
\begin{eqnarray}
H_M^{\ \ N}&=&E_M^{\ \ \Lambda } \delta E_\Lambda^{\ \ N} \\
\Phi_{MN}^{\ \ \ \ P} &=& E_M^{\ \ \Lambda } \delta 
\Omega_{\Lambda N}^{\ \ \ \ P}
\end{eqnarray}
Symmetries that are manifest in superspace are general supercoordinate 
transformations (which include $x$-space diffeomorphisms and local 
supersymmetry), with para\-meters $\xi^\Lambda$, and tangent space (structure 
group) transformations, with parameters $\Lambda^{MN}$. One can solve for 
$H_M^{\ \ N}$ and $\Phi_{MN}^{\ \ \ \ P}$ in terms of these parameters, 
torsions and curvatures as
\begin{eqnarray}
H_M^{\ N}&=& \xi^P T_{PM}^{\ \ \ N} + \nabla_M \xi^N +\Lambda_M^{\ N} \\
\Phi_{MN}^{\ \ \ \ P} &=&\xi^Q R_{QMN}^{\ \ \ \ \ \ P} - \nabla_M
\Lambda_N^{\ P}
\end{eqnarray}
but this does not fix all the degrees of freedom of $H_M^{\ \ N}$ 
\cite{howe82, muller89}. Namely, $H=-\frac{1}{4} H_m^{\ m}$ remains an 
unconstrained superfield and parametrizes the super-Weyl transformations, which
include the dilatations and the special supersymmetry transformations.

${\cal N}=1,2$ Poincar\'e supergravities can be obtained from the 
corresponding conformal theories by consistent couplings to compensating 
multiplets that break superconformal invariance and local U($\cal N$). There 
are different possible choices of compensating multiplets, leading to different
formulations of the Poincar\'e theory. Because of its relevance to this 
paper, we will briefly review the ${\cal N}=2$ case, which was first studied
in \cite{gates80}.

The ${\cal N}=2$ Weyl multiplet has 24+24 degrees of freedom. Its field 
content is given by the graviton $e_{\mu}^m$, the gravitinos $\psi_\mu^{A a}$,
the U(2) connection $\widetilde{\Phi}_\mu^{ab}$, an antisymmetric tensor 
$W_{mn}$ which we decompose as $W_{A \dot A B \dot B} = 2 
\varepsilon_{\dot A \dot B} W_{AB} +2 \varepsilon_{AB} W_{\dot A \dot B}$, a 
spinor $\Lambda_A^a$ and, as auxiliary field, a dimension 2 scalar $I$. In 
superspace, a 
gauge choice can be made (in the supercoordinate transformation) such that the 
graviton and the gravitinos are related to $\theta=0$ components of the 
supervielbein (symbolically $\left. E_\Pi^{\ N }\right| $):
\begin{equation}
\left. E_\Pi ^{\ \ N}\right| =\left[ \begin{array}{ccc} e_\mu^{\ m} &
\frac{1}{2}\psi_\mu^{\ Aa } & \frac{1}{2}\psi_\mu^{\ \dot A a}\\ 
0 & -\delta_B^{\ A} \delta_b^{\ a} & 0 \\ 
0 & 0 & -\delta_{\dot B}^{\ \dot A} \delta_b^{\ a}
\end{array} \right] \label{vielbeinx2}
\end{equation}
In the same way, we gauge the fermionic part of the Lorentz superconnection at 
order $\theta=0$ to zero and we can set its bosonic part equal to the usual 
spin connection: 
\begin{eqnarray}
\left. \Omega_{\mu m}^{\ \ \ n} \right| &=& \omega_{\mu m}^{\ \ \ n}
\left( x\right) \nonumber \\
\left. \Omega_{Aa m}^{\ \ \ \ n}\right|, \left. \Omega_{\dot Aa m}^{\ \ \ \ n}
\right| &=&0 \label{connectionx2}
\end{eqnarray}
The U(2) superconnection $\widetilde{\Phi}_\Pi^{ab}$ is
such that 
\begin{equation}
\left. \widetilde{\Phi}_\mu^{ab}\right|= \widetilde{\Phi}_\mu^{ab}
\label{connectionux2}
\end{equation}
The other fields are the $\theta=0$ component of some superfield, which we 
write in the same way.

The chiral superfield $W_{AB}$ is the basic object of ${\cal N}=2$ 
conformal supergravity, in terms of which its action is written. Other theories
with different ${\cal N}$ have its analogous superfield (e.g. $W_{ABC}$ in 
${\cal N}=1$), but with different spinor and (S)U$\left({\cal N}\right)$ 
indices. A common feature to these superfields is having the antiself-dual
part of the Weyl tensor ${\cal W}_{ABCD}:=-\frac{1}{8} 
{\cal W}^+_{\mu \nu \rho \sigma} \sigma^{\mu \nu}_{\underline{AB}} 
\sigma^{\rho \sigma}_{\underline{CD}}$ in their $\theta$ expansion.

In U(2) ${\cal N}=2$ superspace there is an off-shell solution to the Bianchi
identities. The torsions and curvatures can be expressed in terms of 
superfields $W_{AB}$, $Y_{AB}$, $U_{A \dot A}^{ab}$, $X_{ab}$, their complex 
conjugates and their covariant derivatives. Of these four superfields, only 
$W_{AB}$ transforms covariantly under super-Weyl transformations \cite{howe82}:
\begin{equation}
\delta W_{AB} = H W_{AB}
\end{equation}
The other three superfields transform non-covariantly; they describe all the
non-Weyl covariant degrees of freedom in $H$, and can be gauged away by a 
convenient (Wess-Zumino) gauge choice.

\begin{eqnarray}
\delta Y_{AB} &=& H Y_{AB} -\frac{1}{4} \left[\nabla_A^a, \nabla_{Ba} \right]
H \\
\delta X_{ab} &=& H X_{ab} +\frac{1}{4} \left[\nabla^{\dot A}_a, 
\nabla_{\dot A b} \right] H \\
\delta U^{ab}_{A \dot A} &=& H U^{ab}_{A \dot A} -\frac{1}{2} 
\left[\nabla_A^a, \nabla_{\dot A}^b \right] H
\end{eqnarray}
Another nice feature of ${\cal N}=2$ superspace is that there exists a chiral
density $\epsilon$ and an antichiral projector, given by \cite{muller89}
\begin{equation}
\nabla^{Aa} \nabla_A^b \left(\nabla_a^B \nabla_{Bb} 
+16 X_{ab} \right) - \nabla^{Aa} \nabla_a^B \left(\nabla_A^b \nabla_{Bb} -16i 
Y_{AB} \right)
\end{equation}
When one acts with this projector on any scalar superfield, one gets an 
antichiral superfield (with the exception of $W_{AB}$, only scalar chiral 
superfields exist in curved ${\cal N}=2$ superspace). It is then possible to 
write chiral actions \cite{muller872}.

\subsection{Degauging U(1)}
\indent

The first step for obtaining the Poincar\'e theory is to couple to the 
conformal theory an abelian vector multiplet (with central charge), described 
by a vector $A_\mu$, a complex scalar, a Lorentz-scalar SU(2) triplet and a 
spinorial SU(2) dublet. The vector $A_\mu$ is the gauge field of central charge
transformations; it corresponds, in superspace, to a 1-form $A_\Pi$ with a 
U(1) gauge invariance (the central charge transformation). This 1-form does not
belong to the superspace geometry. 

Using the U(1) gauge invariance we can set the gauge
\begin{equation}
\left. A_\Pi \right| = \left(A_\mu, 0 \right)
\end{equation}
The field strength $F_{\Pi \Sigma}$ is a two-form satisfying its own Bianchi 
identities $\nabla_{\left[\Gamma \right.} F_{\left.\Pi \Sigma \right \}}=0$.
Here we split the U(2) superconnection $\widetilde{\Phi}_\Pi^{ab}$ into a 
SU(2) superconnection $\Phi_\Pi^{ab}$ and a U(1) superconnection 
$\varphi_\Pi$; only the later acts on $A_\Pi$:
\begin{equation}
\widetilde{\Phi}_\Pi^{ab}=\Phi_\Pi^{ab} -\frac{1}{2} \varepsilon^{ab} 
\varphi_\Pi
\end{equation}
One has to impose covariant constraints on its components (like in the 
torsions), in order to construct invariant actions:
\begin{eqnarray}
F_{AB}^{ab}&=& 2 \sqrt{2} \varepsilon_{AB} \varepsilon^{ab} F \\
F_{A \dot B}^{ab}&=&0
\end{eqnarray}
By solving the $F_{\Pi \Sigma}$ Bianchi identities with these constraints, we 
conclude that they define an off-shell ${\cal N}=2$ vector multiplet, given by 
the $\theta=0$ components of the superfields
\begin{equation}
A_\mu, F, F^a_A=\frac{i}{2} F^{\dot A a}_{\ \ \ A \dot A}, F^a_b=\frac{1}{2}
\left(-\nabla^B_b F^a_B +F \overline{X}^a_{\ b}+ \overline{F} X^a_{\ b} \right)
\end{equation}
$\left. F^a_b \right|$ is an auxiliary field; $F^a_a=0$ if the multiplet is 
abelian (as it has to be in this context). $\overline{F}$ is a 
Weyl covariant chiral superfield, with nonzero U(1) and Weyl weigths. A 
superconformal chiral lagrangian for the vector multiplet is given by
\begin{equation}
{\cal L}= \int \overline{\epsilon} F^2 d^4 \overline{\theta} + \mathrm{h.c.}
\end{equation}
In order to get a Poincar\'e theory, we must break the superconformal and local
abelian (from the U(1) subgroup of U(2) - not the gauge invariance of $A_\mu$)
invariances. For that, we set the Poincar\'e gauge
\begin{equation}
F=\overline{F}=1
\end{equation}
As a consequence, from the Bianchi and Ricci identities we get 
\begin{eqnarray}
\varphi_A^a&=&0 \\
F_a^A &=& 0
\end{eqnarray}
Furthermore, $U_{A \dot A}^{ab}$ is an SU(2) singlet, to be identified with the
bosonic U(1) connection (now an auxiliary field):
\begin{equation}
U_{A \dot A}^{ab} = \varepsilon^{ab} U_{A \dot A} = \varepsilon^{ab} 
\varphi_{A \dot A}
\end{equation}\
Other consequences are
\begin{eqnarray}
F_{A \dot A B \dot B} &=& \sqrt{2} i \left[ \varepsilon_{AB} \left(W_{\dot A
\dot B} + Y_{\dot A \dot B} \right) + \varepsilon_{\dot A \dot B}
\left(W_{AB} + Y_{AB} \right) \right] \label{fyw} \\
F^a_b &=& X^a_b \label{fx}\\
\overline{X_{ab}} &=&X^{ab}
\end{eqnarray}
(\ref{fyw}) shows that $W_{mn}$ is now related to the vector field 
strength $F_{mn}$. $Y_{mn}$ emerges as an auxiliary field, like $X_{ab}$
(from (\ref{fx})). We have, therefore, the minimal field representation of
${\cal N}=2$ Poincar\'e supergravity, with a local SU(2) gauge symmetry and 
32+32 off-shell degrees of freedom:
\begin{equation}
e_{\mu}^m, \psi_\mu^{A a}, A_\mu, \Phi_\mu^{ab}, Y_{mn}, U_m, \Lambda_A^a, 
X_{ab}, I \label{32i}
\end{equation}
Although the algebra closes with this multiplet, it does not admit a consistent
lagrangian because of the higher-dimensional scalar $I$ \cite{bs81}.

\subsection{Degauging SU(2)}
\indent

The second step is to break the remaining local SU(2) invariance. This symmetry
can be partially broken (at most, to local SO(2)) through coupling to a 
compensating so-called "improved tensor multiplet" \cite{wpp83, muller871}, 
or broken 
completely. In this work, we take the later possibility. There are still two 
different versions of off-shell ${\cal N}=2$ supergravity without SO(2) 
symmetry, each with different physical degrees of freedom. In both cases we 
start by imposing a constraint on the SU(2) parameter $L^{ab}$ which 
restricts it to a compensating nonlinear multiplet \cite{whp80}
\footnote{Actually, condition (\ref{tensor}) restricts $L^{ab}$ to a tensor
multiplet, which is the linearization of the nonlinear multiplet. This is 
enough for our analysis.}:
\begin{equation}
\nabla_A^{\underline{a}} L^{\underline{b} \underline{c}}=0 \label{tensor}
\end{equation}
From the transformation law of the SU(2) connection
\begin{equation}
\delta \Phi_M^{ab}=-\nabla_M L^{ab}
\end{equation}
we can get the required condition for $L^{ab}$ by imposing the following
constraint on the fermionic connection:
\begin{equation}
\Phi_A^{abc}=2 \varepsilon^{a \underline{b}} \rho_A^{\underline{c}} 
\label{defrho}
\end{equation}
This constraint requires introducing a new fermionic superfield $\rho_A^a$. We
also introduce its fermionic derivatives $P$ (a complex scalar) and $H_m$ (see 
appendix \ref{appendix24}). The previous SU(2) connection $\Phi_\mu^{ab}$
is now an unconstrained auxiliary field. The divergence of the vector field 
$H_m$ is constrained, though, at the linearized level by the condition 
$\nabla^m H_m=\frac{1}{3} R-\frac{1}{12} I$. The full nonlinear constraint is  
\begin{eqnarray}
I &=& 4 R -6 \nabla_{A \dot A} H^{A \dot A} -24 X^{ab} X_{ab} 
-12 W^{AB} Y_{AB} -12 W^{\dot A \dot B} Y_{\dot A \dot B} \nonumber \\
&+& 3 P \overline{P} + \frac{3}{2} H^{A \dot A} H_{A \dot A} 
-12 \Phi^{A \dot A}_{ab} \Phi_{A \dot A}^{ab} -12 U^{A \dot A} U_{A \dot A}
+ 16i \rho^a_A \Lambda^A_a \nonumber \\
&-& 16i \rho^a_{\dot A} \Lambda^{\dot A}_a
-48 \rho^A_a W_{AB}^{\ \ \ Ba} +48 \rho^{\dot A}_a 
W_{\dot A \dot B}^{\ \ \ \dot B a} +48i \rho^A_a \rho^{B a} W_{AB} \nonumber \\
&+& 48i \rho^{\dot A}_a \rho^{\dot B a} W_{\dot A \dot B}
+48 \rho^{A a} \rho^{\dot A}_a U_{A \dot A} -48i \rho^{A a} 
\nabla_{A \dot A} \rho^{\dot A}_a \nonumber \\
&+& 48i \rho^{\dot A a} \nabla_{A \dot A} \rho^A_a +96i \rho^A_a 
\Phi_{A \dot A}^{ab} \rho^{\dot A}_b \label{i}
\end{eqnarray}
which is equivalent to saying that $I$, now defined by (\ref{defi}), is no 
longer an independent field. This constraint implies that only the longitudinal
part of $H_m$ belongs to the nonlinear multiplet; its divergence lies in the
original Weyl multiplet.
From the structure equation 
\begin{equation}
R_{MN}^{\ \ \ \ ab}=E_M^{\ \ \Lambda }E_N^{\ \ \Pi }\left\{ \partial
_\Lambda \Omega_\Pi^{\ ab}+\Omega _\Lambda^{\ ac}\Omega_{\Pi c}^{\ \
\ b}-\left( -\right)^{\Lambda \Pi }\left( \Lambda \leftrightarrow \Pi
\right) \right\}
\end{equation}
and the constraint/definition (\ref{defrho}), we can derive off-shell relations
for the (still SU(2) covariant) derivatives of $\rho^a_A$, which we collect in 
appendix \ref{appendix25}.

Altogether, these component fields form then the "old minimal" ${\cal N}=2$ 
40+40 multiplet \cite{fv79}: 
\begin{equation}
e_{\mu}^m, \psi_\mu^{A a}, A_\mu, \Phi_\mu^{ab}, Y_{mn}, U_m, \Lambda_A^a, 
X_{ab}, H_m, P, \rho_A^a
\end{equation}
This is the formulation of ${\cal N}=2$ supergravity we are working with. 
The other possibility (also with SU(2) completely broken) is to further 
restrict the compensating non-linear multiplet to an on-shell scalar multiplet
\cite{muller86}. This reduction generates a minimal 32+32 multiplet (not to be 
confused with (\ref{32i})) with new physical degrees of freedom. We will not 
pursue this version of ${\cal N}=2$ supergravity in this work.

\subsection{${\cal N}=2$ Poincar\'e supergravity in superspace}
\indent

The final lagrangian of "old minimal" ${\cal N}=2$ supergravity is given by
\footnote{${\cal D}_\mu$ is just the usual Lorentz covariant derivative (not
U(2) covariant).} \cite{whp80,muller84}
\begin{eqnarray}
\kappa^2 {\cal L}_{SG} &=& -\frac{1}{2} e {\cal R} -\frac{1}{4} e 
\varepsilon^{\mu \nu \rho \lambda} \left( \psi_{\mu \dot A}^{\ a} 
\sigma_\nu^{A \dot A} \psi_{\rho \lambda Aa} +\psi_{\mu A}^{\ a}
\sigma_\nu^{A \dot A} \psi_{\rho \lambda \dot Aa} \right) -\frac{1}{4} e
F_{\mu \nu} F^{\mu \nu} \nonumber \\
&-& \frac{1}{4 \sqrt{2}} e \left(\psi_{\mu}^{Aa} \psi_{\nu Aa} +
\psi_{\mu}^{\dot Aa} \psi_{\nu \dot Aa} \right) F^{\mu \nu}
- \frac{i}{4 \sqrt{2}} e \varepsilon^{\mu \nu \rho \lambda} 
\left(\psi_{\mu}^{Aa} \psi_{\nu Aa} - \psi_{\mu}^{\dot Aa} 
\psi_{\nu \dot Aa} \right) F_{\rho \lambda} \nonumber \\
&-& \frac{i}{16} e \varepsilon^{\mu \nu \rho \lambda} 
\left(\psi_{\mu}^{Aa} \psi_{\nu Aa} - \psi_{\mu}^{\dot Aa} 
\psi_{\nu \dot Aa} \right) \left(\psi_{\rho}^{Bb} \psi_{\lambda Bb} +
\psi_{\rho}^{\dot Bb} \psi_{\lambda \dot Bb} \right) \nonumber \\
&-& \frac{1}{4} e Y_{\mu \nu} Y^{\mu \nu} -\frac{1}{2} e X^{ab} X_{ab} +eU^2
-\frac{1}{8} e P \overline{P} -\frac{1}{8} e H^2 +e \Phi^\mu_{ab} \Phi_\mu^{ab}
+\frac{2}{3} ie \rho^{Aa} \Lambda_{Aa} \nonumber \\
&-&\frac{2}{3} ie \rho^{\dot Aa} \Lambda_{\dot Aa} 
-4e \rho^{Aa} W_{ABa}^{\ \ \ B} +4e \rho^{\dot Aa} 
W_{\dot A \dot Ba}^{\ \ \ \dot B} +2ie\rho^{Aa} \rho^B_a W_{AB} 
+2ie\rho^{\dot Aa} \rho^{\dot B}_a W_{\dot A \dot B} \nonumber \\
&-&ie P \rho^{Aa} \rho_{Aa} -ie \overline{P} \rho^{\dot Aa} \rho_{\dot Aa}
-2e\rho^{Aa} \rho^{\dot A}_a U_{A \dot A}  \nonumber \\
&+&2i e \rho^{Aa} \sigma^\mu_{A \dot A} {\cal D}_\mu \rho^{\dot A}_a -2i e 
\rho^{\dot Aa} \sigma^\mu_{A \dot A} {\cal D}_\mu \rho^A_a +ie \left(
\rho^{Aa} \sigma^\mu_{A \dot A} \psi_{\mu}^{\dot A b} +
\rho^{\dot Aa} \sigma^\mu_{A \dot A} \psi_\mu^{A b} \right) X_{ab}\nonumber \\
&+& e \rho^{Aa} \sigma^\mu_{A \dot A} \psi_{\mu \dot B a} Y^{\dot A \dot B}
+ e \rho^{\dot Aa} \sigma^\mu_{A \dot A} \psi_{\mu B a} Y^{AB}
+ ie \rho^{Aa} \sigma^\mu_{A \dot A} \psi_{\mu B a} U^{B \dot A} \nonumber \\
&+& ie \rho^{\dot Aa} \sigma^\mu_{A \dot A} \psi_{\mu \dot B a} U^{A \dot B}
+ \frac{1}{4} e \rho^{Aa} \sigma^\mu_{A \dot A} \psi_{\mu a}^{\ \dot A} P
+ \frac{1}{4} e \rho^{\dot Aa} \sigma^\mu_{A \dot A} \psi^{\ A}_{\mu a}
\overline{P} \nonumber \\
&-& \left(\rho^{Aa} \psi^{\ \ b}_{\mu A}-
\rho^{\dot A a} \psi^{\ \ b}_{\mu \dot A} \right) \left(\rho^B_b 
\sigma^{\mu \nu}_{BC} \psi_{\nu a}^C - \rho^{\dot B}_b 
\sigma^{\mu \nu}_{\dot B \dot C} \psi_{\nu a}^{\dot C} \right) 
-\rho^{\dot A}_a \sigma_\mu^{A \dot B} \psi^{\mu}_{\dot B b} 
\Phi_{A \dot A}^{ab} \nonumber \\
&+& \rho^A_a \sigma_\mu^{B \dot A} \psi^{\mu}_{B b} \Phi_{A \dot A}^{ab}
- \frac{1}{4} e \rho^{Aa} \sigma_\mu^{B \dot A} \psi^{\mu}_{B a} H_{A \dot A}
+ \frac{1}{4} e \rho^{\dot Aa} \sigma_\mu^{A \dot B} \psi^{\mu}_{\dot B a} 
H_{A \dot A}  
\end{eqnarray}

The final solutions to the Bianchi identities in SU(2) ${\cal N}=2$ superspace
are listed in appendix \ref{appendix2}. We present both the expressions for 
the torsions and curvatures and the off-shell 
differential relations among the superfields (appendix \ref{appendix25}). As 
first noticed in \cite{gates80}, these solutions only depend on $W_{AB}$ (a 
physical field at $\theta=0$), $\rho_A^a$ (an auxiliary field at $\theta=0$), 
their complex conjugates and their covariant derivatives. Here we present for 
completeness the full expansion of the (anti)chiral density 
$\overline{\epsilon}$ \cite{muller84}:

\begin{eqnarray}
\overline{\epsilon}&=&e -i e \theta^{\dot A}_a \sigma^\mu_{A \dot A} 
\psi^{Aa}_\mu \nonumber \\
&+& \frac{e}{2} \theta^{\dot A}_a \theta^{\dot B}_b \left[\frac{16}{3}
\varepsilon_{\dot A \dot B} X^{ab} +\frac{8}{3}i \varepsilon^{ab} 
Y_{\dot A \dot B} +\sigma^\mu_{A \dot A} \psi^{Aa}_\mu \sigma^\nu_{B \dot B} 
\psi^{Bb}_\nu - \sigma^\mu_{A \dot A} \psi^{Aa}_\nu \sigma^\nu_{B \dot B} 
\psi^{Bb}_\mu \right] \nonumber \\
&-& \frac{e}{6} \theta^{\dot A}_a \theta^{\dot B}_b \theta^{\dot C}_c 
\left[\frac{16}{3}i \varepsilon_{\dot A \dot B} \varepsilon^{ac}
\Lambda_{\dot C}^b +4 \varepsilon_{\dot A \dot B} \varepsilon^{ac}
\sigma^\mu_{C \dot C} \psi_{\mu B}^b Y^{BC} \right. \nonumber \\
&-&i \varepsilon_{\dot A \dot B} \varepsilon^{ac}
\sigma^\mu_{C \dot C} \sigma_{\rho \lambda}^{BC} \psi^{\ b}_{\mu B}
\left(\sqrt{2} F^{\rho \lambda} + \psi^{\rho Ee} \psi_{Ee}^\lambda 
+\psi^{\rho \dot Ee} \psi_{\dot Ee}^\lambda \right) \nonumber \\
&+& 2 \sigma^\mu_{A \dot A} \psi^{Aa}_\mu \left(7i \varepsilon_{\dot B \dot C}
X^{bc} -5 \varepsilon^{bc} Y_{\dot B \dot C} \right) +i \sigma^\mu_{A \dot A} 
\psi^{Aa}_\mu \sigma^\nu_{B \dot B} \psi^{Bb}_\nu 
\sigma^\lambda_{C \dot C} \psi^{Cc}_\lambda \nonumber \\
&-& \left. 3i\sigma^\mu_{A \dot A} \psi^{Aa}_\mu \sigma^\nu_{B \dot B} 
\psi^{Bb}_\lambda \sigma^\lambda_{C \dot C} \psi^{Cc}_\nu
+2i \sigma^\mu_{A \dot A} \psi^{Aa}_\nu \sigma^\nu_{B \dot B} \psi^{Bb}_\lambda
\sigma^\lambda_{C \dot C} \psi^{Cc}_\mu \right]  \nonumber \\
&-& \frac{2}{3} \theta^{\dot A}_a \theta^{\dot B}_b \theta_{\dot A}^b 
\theta_{\dot B}^a \left[\partial_\mu \left(-i e U^\mu + \frac{e}{2} H^\mu 
+2e \rho^A_a \sigma^{\mu \nu}_{A B} \psi_\nu^{B a} -2e \rho^{\dot A}_a 
\sigma^{\mu \nu}_{\dot A \dot B} \psi_\nu^{\dot B a} \right) \right.
\nonumber \\
&+&\frac{e}{8} \varepsilon^{\mu \nu \rho \lambda}  \left(i F_{\mu \nu} 
F_{\rho \lambda} - \psi^{\ \ a}_{\mu A} \sigma_\nu^{A \dot A}
\psi_{\rho \lambda \dot A a} +\psi^{\ \ a}_{\mu \dot A} \sigma_\nu^{A \dot A}
\psi_{\rho \lambda A a}  \right. \nonumber \\
&-& \left. \left. i \psi_{\mu}^{Aa} \psi_{\nu Aa} \psi_\rho^{\dot A b} 
\psi_{\lambda \dot A b}  \right) + \kappa^2 {\cal L}_{SG} \right] 
\label{chiral2}
\end{eqnarray}
This allows us to write, up to total derivatives,
\begin{equation}
{\cal L}_{SG} = -\frac{3}{4 \kappa^2} \int \overline{\epsilon}
d^4 \overline{\theta} + \mathrm{h.c.}
\end{equation}

\section{The supersymmetric ${\cal R}^4$ lagrangian}
\setcounter{equation}{0}
\indent

Our goal in this article is to supersymmetrize the fourth power of the Weyl 
tensor. (There are indeed, in four dimensions, thirteen independent scalar 
fourth-degree polynomials of the Riemann tensor, but we are only interested in
a particular one. See \cite{moura02} for a complete discussion.) As mentioned 
before, $W_{AB}$ contains in its $\theta$ expansion the antiself-dual part of 
the Weyl tensor (see (\ref{w41})).


\subsection{The lagrangian in superspace}
\indent

Analogously to \cite{moura02}, we write the supersymmetric ${\cal R}^4$ 
lagrangian in superspace, using the chiral projector and the chiral density, as
a (quantum) correction to the pure supergravity lagrangian:
\begin{eqnarray}
{\cal L}&=& \int \overline{\epsilon} \left[-\frac{3}{4 \kappa^2} +\alpha
\kappa^4 \left(\nabla^{Aa} \nabla_A^b \left(\nabla_a^B \nabla_{Bb} 
+16 X_{ab} \right) \right. \right. \nonumber \\
&-& \left. \left. \nabla^{Aa} \nabla_a^B \left(\nabla_A^b \nabla_{Bb} -16i 
Y_{AB} \right) \right)
W^2\overline{W}^2 \right] d^4 \overline{\theta} + \mathrm{h.c.} \nonumber \\
&=&{\cal L}_{SG} + {\cal L}_{{\cal R}^4} \label{action2}
\end{eqnarray}
$\alpha$ is a (numerical) constant (we use a different definition from 
\cite{moura02,moura01}). Up to that (unknown) numerical factor, this can be 
seen as a three-loop ${\cal N}=2$ supergravity effective action or, 
equivalently, as a four dimensional ${\cal N}=2$ string/M-theory effective 
action resulting from a compactification and truncation from ten/eleven 
dimensions.

After the superspace integration, (\ref{action2}) comes, in terms of 
components, as
\begin{eqnarray}
{\cal L}_{{\cal R}^4}&=& \alpha \kappa^4 \int \epsilon \phi d^4 \theta + 
\mathrm{h.c.} \nonumber \\
&=& \alpha \kappa^4 \left(\epsilon^{(0)} \phi^{(4)} +\frac{1}{36} 
\epsilon^{(1)Bb} \phi^{(3)aA}_{B\ \ \ bAa} +\frac{1}{48} 
\epsilon^{(2)bAa}_A \phi^{(2)B}_{\ \ \ \ aBb} \right. \nonumber \\
&-&\left. \frac{1}{48} \epsilon^{(2)\ \ a}_{BaA}
\phi^{(2)AbB}_{\ \ \ \ \ \ \ b} +\frac{1}{36} \phi^{(1)Bb} 
\epsilon^{(3)aA}_{B\ \ \ bAa} + \epsilon^{(4)} \phi^{(0)} \right)+ 
\mathrm{h.c.} \label{action2chiral}
\end{eqnarray}
where we have defined the chiral superfield
\begin{equation}
\phi=\left(\nabla^{\dot Aa} \nabla_{\dot A}^b \left(\nabla_a^{\dot B} 
\nabla_{\dot Bb} +16 X_{ab} \right)  
-\nabla^{\dot Aa} \nabla_a^{\dot B} \left(\nabla_{\dot A}^b \nabla_{\dot Bb} 
-16i Y_{\dot A \dot B} \right) \right) W^2\overline{W}^2 \label{defphi}
\end{equation}
and
\begin{eqnarray}
\phi^{(0)}&=& \left. \phi \right| \nonumber \\
\phi^{(1)Aa}&=& \left. \nabla^{Aa} \phi \right|\nonumber \\
\phi^{(2)}_{AaBb} &=& \frac{1}{2} \left. \left[\nabla_{Aa}, \nabla_{Bb} 
\right] \phi \right|\nonumber \\
\phi^{(3)}_{AaBbCc}&=& \frac{1}{6} \left(\nabla_{Aa} \left[\nabla_{Bb}, 
\nabla_{Cc} \right]+ \nabla_{Bb} \left[\nabla_{Cc}, \nabla_{Aa} \right]
+\nabla_{Cc} \left[\nabla_{Aa}, \nabla_{Bb} \right] \right) 
\left. \phi \right|\nonumber \\
\phi^{(4)}&=& \frac{1}{288} \left. \nabla^{Aa} \nabla^{Bb}
\left[\nabla_{Ab}, \nabla_{Ba} \right]\phi \right| \label{components}
\end{eqnarray}
As one can see from the component expansions in appendix \ref{appendix2}, the 
term $\epsilon^{(0)} \phi^{(4)} +\mathrm{h.c.}$ clearly contains the fourth 
power of the Weyl tensor, more precisely (using the notation of \cite{moura02})
$e {\cal W}_+^2 {\cal W}_-^2$. 


\subsection{The lagrangian in components}
\indent

We now proceed with the calculation of the components of $\phi$ and analysis of
its field content. For that, we use the
differential constraints from the solution to the Bianchi identities and the 
commutation relations listed in appendix \ref{appendix2} to compute the
components in (\ref{components}). The process is straightforward but lengthy.

We start by expanding $\phi$ as
\begin{eqnarray}
\phi &=& W^2 \nabla^{\dot Aa} \nabla_{\dot A}^b 
\nabla_a^{\dot B} \nabla_{\dot Bb} \overline{W}^2 - W^2 \nabla^{\dot Aa} 
\nabla_a^{\dot B} \nabla_{\dot A}^b \nabla_{\dot Bb} \overline{W}^2\nonumber \\
&-& 32 i W^2 \overline{W}^2 \nabla^{\dot Aa} \Lambda_{\dot Aa} +32i
W^2 \Lambda^{\dot Aa} \nabla_{\dot Aa} \overline{W}^2 \nonumber \\
&-& 64i W^2 \left(\nabla^{\dot Aa} X_a^{\ b} \right) W^{\dot B \dot C}
W_{\dot B \dot C \dot A b}
- 64i W^2  X_a^{\ b} \left(\nabla^{\dot Aa} W^{\dot B \dot C} \right)
W_{\dot B \dot C \dot A b} \nonumber \\
&-& 64i W^2  X_a^{\ b} W^{\dot B \dot C} \nabla^{\dot Aa} 
W_{\dot B \dot C \dot A b} - \frac{64}{3} W^2 \left( \nabla^{\dot Aa} X_{ab}
\right) W_{\dot A \dot B} \Lambda^{\dot Bb} \nonumber \\
&-& \frac{64}{3} W^2 X_{ab} \left(\nabla^{\dot Aa} W_{\dot A \dot B} \right)
\Lambda^{\dot Bb} - \frac{64}{3} W^2  X_{ab} W_{\dot A \dot B} \nabla^{\dot Aa}
\Lambda^{\dot Bb} \nonumber \\
&-& 64 W^2 \left(\nabla_{\dot A}^a Y^{\dot A \dot B} \right) W^{\dot C \dot D}
W_ {\dot C \dot D \dot B a} -64 W^2 Y^{\dot A \dot B} \left( \nabla_{\dot A}^a
W^{\dot C \dot D} \right) W_ {\dot C \dot D \dot B a} \nonumber \\
&-& 64 W^2 Y^{\dot A \dot B} W^{\dot C \dot D} \nabla_{\dot A}^a 
W_ {\dot C \dot D \dot B a} +\frac{64}{3}i W^2 \left(\nabla^{\dot Aa} 
Y_{\dot A \dot B} \right) W^{\dot B \dot C} \Lambda_{\dot Ca} \nonumber \\
&+& \frac{64}{3}i W^2 Y_{\dot A \dot B} \left(\nabla^{\dot Aa} 
W^{\dot B \dot C} \right) \Lambda_{\dot Ca} +\frac{64}{3}i W^2 
Y_{\dot A \dot B} W^{\dot B \dot C} \nabla^{\dot Aa} \Lambda_{\dot Ca}
\label{phiexp}
\end{eqnarray}
All the terms in (\ref{phiexp}) can be immediately computed by using the 
differential relations in appendix \ref{appendix25}, except for the first two,
which require a substantial amount of derivatives to compute. 

In order to compute $\nabla_{\dot Aa} \overline{W}^2$, we use the known 
relation for $\nabla_{\dot Aa} W_{\dot C \dot D}$:
\begin{equation}
\nabla_{\dot Aa} \overline{W}^2 = 2 W^{\dot C \dot D} \nabla_{\dot Aa} 
W_{\dot C \dot D} = 4i W^{\dot C \dot D} W_{\dot C \dot D \dot Aa}
- \frac{4}{3} W_{\dot A \dot B} \Lambda^{\dot B}_a
\end{equation}
With this result, we can compute
\begin{eqnarray}
\nabla_{\dot Bb} \nabla_{\dot Aa} \overline{W}^2 &=& 4i \left(\nabla_{\dot Bb}
W^{\dot C \dot D} \right) W_ {\dot C \dot D \dot A a} +4i W^{\dot C \dot D}
\nabla_{\dot Bb} W_ {\dot C \dot D \dot A a} \nonumber \\
&-& \frac{4}{3} \left(\nabla_{\dot Bb} W_{\dot A \dot C} \right) 
\Lambda^{\dot C}_a - \frac{4}{3} W_{\dot A \dot C} \nabla_{\dot Bb}
\Lambda^{\dot C}_a \label{d2w2}
\end{eqnarray}
The field content implicit in these relations may be seen from the differential
relations in appendix \ref{appendix25}.

From (\ref{d2w2}), we can proceed computing:
\begin{eqnarray}
\nabla_{\dot Cc} \nabla_{\dot Bb} \nabla_{\dot Aa} \overline{W}^2 &=&
4i \left(\nabla_{\dot Cc} \nabla_{\dot Bb} W^{\dot D \dot E} \right) 
W_{\dot D \dot E \dot A a} -4i \left(\nabla_{\dot Bb} W^{\dot D \dot E} 
\right) \nabla_{\dot Cc} W_{\dot D \dot E \dot A a} \nonumber \\
&+&4i \left(\nabla_{\dot Cc} W^{\dot D \dot E} \right) 
\nabla_{\dot Bb} W_{\dot D \dot E \dot A a} +4i W^{\dot D \dot E}
\nabla_{\dot Cc} \nabla_{\dot Bb} W_{\dot D \dot E \dot A a} \nonumber \\
&-& \frac{4}{3} \left(\nabla_{\dot Cc} \nabla_{\dot Bb} W_{\dot A \dot D} 
\right) \Lambda^{\dot D}_a - \frac{4}{3} \left(\nabla_{\dot Cc} 
W_{\dot A \dot C} \right) \nabla_{\dot Bb} \Lambda^{\dot C}_a \nonumber \\
&-& \frac{4}{3} \left(\nabla_{\dot Cc} W_{\dot A \dot D} \right) 
\nabla_{\dot Bb} \Lambda^{\dot D}_a - \frac{4}{3} W_{\dot A \dot D} 
\nabla_{\dot Cc} \nabla_{\dot Bb} \Lambda^{\dot D}_a \label{d3w2}
\end{eqnarray}
We got some second spinorial derivatives of superfields, as expected, which we 
can compute by differentiating some of the relations in appendix 
\ref{appendix25}. We list the results in appendix \ref{appendix31}.

From (\ref{d3w2}), we finally get 
\begin{eqnarray}
\nabla_{\dot Dd} \nabla_{\dot Cc} \nabla_{\dot Bb} \nabla_{\dot Aa} 
\overline{W}^2 =
-\frac{4}{3} \left(\nabla_{\dot Dd} \nabla_{\dot Cc} \nabla_{\dot Bb} 
W_{\dot A \dot E} \right) \Lambda^{\dot E}_a &-&\frac{4}{3} 
\left(\nabla_{\dot Cc} \nabla_{\dot Bb} W_{\dot A \dot E} \right) 
\nabla_{\dot Dd} \Lambda^{\dot E}_a \nonumber \\
+\frac{4}{3} \left(\nabla_{\dot Dd} \nabla_{\dot Bb} W_{\dot A \dot E} 
\right) \nabla_{\dot Cc} \Lambda^{\dot E}_a
&-&\frac{4}{3} \left(\nabla_{\dot Bb} W_{\dot A \dot E} \right) 
\nabla_{\dot Dd} \nabla_{\dot Cc} \Lambda^{\dot E}_a \nonumber \\
-\frac{4}{3} \left(\nabla_{\dot Dd} \nabla_{\dot Cc} W_{\dot A \dot E} 
\right) \nabla_{\dot Bb} \Lambda^{\dot E}_a 
&+&\frac{4}{3} \left(\nabla_{\dot Cc}
W_{\dot A \dot E} \right) \nabla_{\dot Dd} \nabla_{\dot Bb} \Lambda^{\dot E}_a 
\nonumber \\
-\frac{4}{3} \left(\nabla_{\dot Dd} W_{\dot A \dot E} \right) \nabla_{\dot Cc} 
\nabla_{\dot Bb} \Lambda^{\dot E}_a &-&\frac{4}{3} W_{\dot A \dot E} 
\nabla_{\dot Dd} \nabla_{\dot Cc} \nabla_{\dot Bb} \Lambda^{\dot E}_a
\nonumber \\
+4i \left(\nabla_{\dot Dd} \nabla_{\dot Cc} \nabla_{\dot Bb} 
W^{\dot E \dot F} \right) W_{\dot E \dot F \dot A a}
&+& 4i \left(\nabla_{\dot Cc} \nabla_{\dot Bb} W^{\dot E \dot F} \right) 
\nabla_{\dot Dd} W_{\dot E \dot F \dot A a} \nonumber \\
-4i \left(\nabla_{\dot Dd} 
\nabla_{\dot Bb} W^{\dot E \dot F} \right) \nabla_{\dot Cc} 
W_{\dot E \dot F \dot A a}
&+&4i \left(\nabla_{\dot Bb} W^{\dot E \dot F} \right) \nabla_{\dot Dd} 
\nabla_{\dot Cc} W_{\dot E \dot F \dot A a} \nonumber \\
+ 4i \left(\nabla_{\dot Dd} \nabla_{\dot Cc} W^{\dot E \dot F} \right) 
\nabla_{\dot Bb} W_{\dot E \dot F \dot A a} &-&4i \left(\nabla_{\dot Cc} 
W^{\dot E \dot F} \right) \nabla_{\dot Dd} \nabla_{\dot Bb} 
W_{\dot E \dot F \dot A a} \nonumber \\
+4i \left(\nabla_{\dot Dd} W^{\dot E \dot F} \right) \nabla_{\dot Cc} 
\nabla_{\dot Bb} W_{\dot E \dot F \dot A a} &+&4i W^{\dot E \dot F} 
\nabla_{\dot Dd} \nabla_{\dot Cc} \nabla_{\dot Bb} W_{\dot E \dot F \dot A a} 
\label{d4w2}
\end{eqnarray}
As expected, we get some third spinorial derivatives of superfields,
which we can compute by differentiating some of the relations from the previous
section and in appendix \ref{appendix25}. We list the results in appendix
\ref{appendix32}.

When a spinor derivative acts on a vector derivative of a superfield, first we 
commute the two derivatives with the help of the torsions and curvatures listed
in the appendix \ref{appendix2}. Then we use the gauge choices 
(\ref{vielbeinx2}), (\ref{connectionx2}) and (\ref{connectionux2}) to write
\begin{equation}
\left. \nabla_\mu \right| = \nabla_\mu
\end{equation}
$\nabla_\mu$ is a Lorentz and SU(2) covariant derivative. For an arbitrary 
superfield $G$ we have then
\begin{equation}
\left. \nabla_m G \right|= e_m^{\ \mu} \nabla_\mu \left. G \right|
-\frac{1}{2} \psi_m^{Aa} \left. \nabla_{Aa} G \right|
-\frac{1}{2} \psi_m^{\dot A a} \left. \nabla_{\dot A a} G \right|
\end{equation}
The equations we have been obtaining allow us to determine the field content of
$\left. \phi \right|$, by replacing the equations with fewer spinorial
derivatives (starting from the differential relations in appendix 
\ref{appendix25}) in the ones with more spinorial derivatives, and by suitable 
index contraction and derivative commutation.

This same relations (more precisely, their complex conjugates) will also be 
useful for calculating the higher $\theta$ components of $\phi$. Other 
differential relations will be ne\-cessary: the series of spinorial derivatives
will act on $\left. \phi \right|$, which after the full component expansion 
contains much more fields (not only $W_{AB}$). These fields will also be acted 
on by four spinorial derivatives, which we have not computed (and, as we will 
see next section, we don't need to). The computations are
straightforward, like the one we did for $\overline{W}^2$. Rather than 
performing them, we prefer to deduce some of their properties, using the 
results we have.
\section{The field equations for the auxiliary fields}
\label{4}
\setcounter{equation}{0}
\indent

Having seen how to obtain the supersymmetric ${\cal R}^4$ lagrangian in 
components, we are now in a position to analyze the auxiliary field sector. 
Our main goal is to figure out which auxiliary fields, with this ${\cal R}^4$ 
correction, do not get spacetime derivatives in the action (i.e. have an 
algebraic field equation and can be eliminated on-shell), and which do get. We 
start by the ${\cal N}=2$ case, the lagrangian of which we have been 
determining. Then we summarize the ${\cal N}=1$ case, which we analyzed in 
previous works, and we compare the two cases.
\subsection{The ${\cal N}=2$ case}
\indent

We start by recalling that, in pure supergravity, both in ${\cal N}=1$ and in 
${\cal N}=2$, the auxiliary fields are equal to 0 on-shell \cite{muller84}.

Just by looking at the differential relations in appendix \ref{appendix25}, it 
is immediate to conclude that, if a $\nabla_{\dot Bb} W_{\dot C\dot D \dot Aa}$
term shows up in the lagrangian, one gets derivatives of $\Phi_{A\dot B}^{ab}$ 
and of $U_m$. This term shows up already in $\nabla_{\dot Bb} 
\nabla_{\dot Aa} \overline{W}^2$ (which shows up, by itself, in higher $\theta$
components of $\overline{\phi}$). Dotted spinor derivatives of 
$\Phi_{A\dot B}^{ab}$ and 
$U_m$ will introduce the physical superfields $W_{\dot C\dot D}, 
W_{\dot C\dot D \dot Aa}, Y_{\dot C\dot D Aa}$ and the auxiliary superfields 
$Y_{\dot C\dot D}, X_{ab}, \Lambda_{Aa}$ (but not $\Lambda_{\dot Aa}$).
It also introduces a derivative of the superfield $\rho_{\dot A a}$.
These superfields get then one derivative - $\rho_{\dot A a}$ actually gets
two - in the term $\nabla_{\dot C c} \nabla_{\dot Bb} \nabla_{\dot Aa} 
\overline{W}^2$.

Just by inspection, we expect these auxiliary fields to have derivatives. To
actually compute the coefficient of their derivative terms is a hard task - 
basically it would be equivalent to computing all the terms in the lagrangian, 
which would require an enormous amount of algebra. Fortunately, it is possible 
to compute their leading derivative terms using a simple trick.

In the previous section we obtained an expression for $\left. \phi \right|$ 
in terms of spinorial derivatives of superfields. To compute higher
$\theta$ terms, we act on $\phi$ with undotted spinor 
derivatives. From (\ref{phiexp}) we see that these derivatives either act on 
$W^2$, $W_{\dot A \dot B \dot C c}$, $X_{ab}$, $\Lambda_{\dot C c}$, 
$Y_{\dot A \dot B}$ or their spinor derivatives - giving rise to the equations 
we got in computing
$\left. \phi \right|$, but complex-conjugated -, or they act on four dotted
spinorial derivatives of $\overline{W}^2$. Each undotted derivative can be 
anticommuted with a dotted one (giving rise to a vector derivative and 
curvature terms), until it finally acts in $\overline{W}^2$ - resulting 0.
The torsion and curvature terms we get in this process, after all undotted 
derivatives are taken, include other superfields like $U_{A \dot A}$, 
$Y_{A B \dot C c}$,
the spinor derivatives of which we know from computing $\left. \phi \right|$.
Therefore, the higher $\theta$ terms of $\phi$ are either products of terms we
already know from $\left. \phi \right|$, or overall vector derivatives. Since 
the higher $\theta$ terms are multiplied, in the action (\ref{action2chiral}), 
by a corresponding term of the chiral density, we can integrate by parts the
derivatives in the higher $\theta$ terms and have them acting in the chiral 
density! Therefore, all the lower $\theta^{4-n}$ terms in the chiral density 
are acted in the action by $n$ vector derivatives. Looking at the chiral 
density in (\ref{chiral2}), we conclude that $X_{ab}$ and $Y_{mn}$ have terms
with at least two derivative in the action, and $\Lambda_{Aa}$ has terms with 
at least one derivative in the action. These terms cannot be eliminated by 
integration by parts.

Furthermore, $U_m$ and $\Phi_m^{ab}$ also get derivatives in the action. The
reason is the following: the vector derivatives we get come from superspace
and, therefore, are (or can be made) Lorentz and U(2) covariant.
\footnote{In this paper, they are just SU(2) covariant, by our choice. The 
choice of tangent group in superspace is a matter of convenience. We could 
have kept the U(2) covariance, and the argument in the text would be valid 
directly, but we broke this covariance to SU(2). We get them extra $U_m$ terms
 - $U_m$ was the U(1) connection -, which would be reabsorbed in the 
derivatives, if the U(2) covariance had been kept.} The physical theory 
should be only Lorentz invariant; the U(1) and SU(2) connections should be seen
respectively as the $U_m$ and $\Phi_m^{ab}$ auxiliary fields. Therefore, terms 
with $n$ vector (Lorentz and U(2) covariant) derivatives give rise to $n-1$
vector Lorentz covariant derivatives of $U_m$ and $\Phi_m^{ab}$.

Derivatives of $P$ and $H_m$ superfields still have not appeared up to now. By 
our previous arguments, if these derivatives do not appear in 
$\left. \phi \right|$, they do not appear at all, since these superfields do 
not appear in the chiral density except in the highest $\theta$ term. 
Therefore, all that is left to do is check that $\left. \phi \right|$ has no 
derivatives of $\rho_A^a$, $P$, their hermitean conjugates and $H_m$.

\subsubsection{Derivatives of $P$ and $H_m$ in $\nabla_{\dot Bb} 
\nabla_{\dot Aa} \overline{W}^2$ and $\nabla_{\dot Cc} 
\nabla_{\dot Bb} \nabla_{\dot Aa} \overline{W}^2$}
\indent

The $P$ and $H_m$ superfields only appear in $\nabla_{\dot Bb} \nabla_{\dot Aa}
\overline{W}^2$ through $I$, which belongs to the original Weyl multiplet and 
is given by (\ref{i}), and its derivatives \footnote{We will be only looking 
for derivatives of $P$ because, although this superfield is chiral, it is 
generated (and acquires derivatives) through $\nabla_{\dot Bb} \rho_{\dot A a}$
and $\nabla_{\dot Bb} H_{A \dot A}$. Its complex conjugate, $\overline{P}$,
does not get derivatives in $\phi$, but gets them in $\overline{\phi}$.}. 
Recall that $I$ itself has a divergence 
$\nabla_{B \dot B} H^{B \dot B}$ and derivatives $\nabla_{B \dot B} 
\rho_{\dot A a}$; it shows up already in the expansion of $\nabla_{\dot Bb} 
\nabla_{\dot Aa} \overline{W}^2$ in (\ref{d2w2}), through $\nabla_{\dot Bb} 
\Lambda^{\dot Bb}$. 
These derivatives already show up at this early stage, and will keep showing up
as the calculation goes. We anticipate that we will be interested only in 
derivatives of fields coming from the nonlinear multiplet. Therefore, from now 
on when we refer to "derivatives of $H_m$" we are excluding its divergence.

As we keep going and compute $\nabla_{\dot Cc} \nabla_{\dot Bb} 
\nabla_{\dot Aa} \overline{W}^2$, we see that the derivatives in which we are 
interested come only from the $\nabla_{\dot Cc} I$ term; all the other terms in
(\ref{d3w2}), which we computed in terms of known expressions in appendix 
\ref{appendix31}, contribute at most with $\nabla_{B \dot B} H^{B \dot B}$ 
terms. 

The complete expansion for $\nabla_{\dot Cc} I$ is given by (\ref{di}) in 
appendix \ref{appendix31}. For each term in this equation, we compute in 
appendix \ref{appendix41} its content in terms of the derivatives of interest. 
Now, here a surprise happens: after all the contributions have been analyzed, 
we conclude that they all precisely cancel in such a way that 
$\nabla_{\dot Cc} I$ contains no derivatives of $P$ or of $H_m$ (with the
exception of $\nabla_{B \dot B} H^{B \dot B}$, as before). This result is 
easily obtained from (\ref{di}) and the expressions from appendix 
\ref{appendix41}. This shows that derivatives of $P$ and $H_m$ do not come up 
to $\nabla_{\dot Cc} \nabla_{\dot Bb} \nabla_{\dot Aa} \overline{W}^2$. 

\subsubsection{Derivatives of $P$ and $H_m$ in $\nabla_{\dot Dd} 
\nabla_{\dot Cc} \nabla_{\dot Bb} \nabla_{\dot Aa} \overline{W}^2$}
\indent

An even bigger surprise is that these derivatives also cancel in those terms 
from $\nabla_{\dot Dd} \nabla_{\dot Cc} \nabla_{\dot Bb} \nabla_{\dot Aa} 
\overline{W}^2$ in which we will be interested. The procedure we took to fully 
check these cancellations was analogous to the previous 
case: we expanded each term in (\ref{d4w2}) in terms of known expressions in 
appendix \ref{appendix32}. For those terms which had derivatives of 
interest, we computed their derivative content in appendix \ref{appendix42}.
We summed all the possible contributions for each term in (\ref{d4w2}) and we 
list them next. During the partial calculations we see lots of independent 
terms appearing from different sources, but at the end we found several 
remarkable cancellations of those terms. The most remarkable one happens in
$\nabla_{\dot Dd} \nabla_{\dot Cc} I$, which is part of the term with
$\nabla_{\dot Dd} \nabla_{\dot Cc} \nabla_{\dot Bb} \Lambda_{\dot Aa}$ in 
(\ref{d4w2}). As it can be seen from the expansion of $\nabla_{\dot Dd} 
\nabla_{\dot Cc} I$ in equation (\ref{ddi}), in order to get the derivative 
content of this term we need contributions from every expression in appendix 
\ref{appendix4}. It turns out, though, after a lot of algebra, that all the 
contributions precisely cancel in such a way
that $\nabla_{\dot Dd} \nabla_{\dot Cc} I$ has absolutely no derivatives 
of $P$ or $H_m$ besides $\nabla_{B \dot B} H^{B \dot B}$, exactly like 
$\nabla_{\dot Cc} I$ in the previous section. All these results can be checked 
using the equations in appendices \ref{appendix3} and \ref{appendix4}. After 
performing all the calculations, the final result is then the following:
the only nonvanishing contributions to interesting derivative terms in 
(\ref{d4w2}) come from
\begin{equation}
\nabla_{\dot Dd} \nabla_{\dot Cc} \nabla_{\dot Bb}
W_{\dot E \dot F \dot Aa}= \frac{1}{4} \varepsilon_{\dot B \dot A} 
\varepsilon_{\dot D \dot C} \varepsilon_{dc} X_{ab}
\nabla^A_{\ \underline{\dot E}} H_{A \underline{\dot F}} +\frac{1}{4} 
\varepsilon_{\dot B \dot A} \varepsilon_{\dot D \dot C} \varepsilon_{dc}
\Phi^A_{\ \underline{\dot E}ab} \nabla_{A \underline{\dot F}} P +\cdots 
\label{ddddw}
\end{equation}
These terms actually do not vanish in $\nabla_{\dot Dd} \nabla_{\dot Cc} 
\nabla_{\dot Bb} \nabla_{\dot Aa} \overline{W}^2$, but they do vanish in 
$\phi$. This is because, as it may be seen from (\ref{phiexp}), $\phi$
only contains the combinations $\nabla^{\dot A \underline{a}} 
\nabla_{\dot A}^{\underline{b}} \nabla_a^{\dot B} \nabla_{\dot Bb} 
\overline{W}^2$ and $\nabla^{\underline{\dot A}a} \nabla_a^{\underline{\dot B}}
\nabla_{\dot A}^b \nabla_{\dot Bb} \overline{W}^2$. In the first combination, 
we need $\nabla^{\dot D}_{\underline{d}} \nabla_{\dot D \underline{c}} 
\nabla^{\dot B c} W_{\dot E \dot F \dot B}^{\ \ \ \ \ d}$, while in the second 
we need $\nabla^{\underline{\dot D} c} \nabla^{\underline{\dot C}}_c 
\nabla_{\dot D}^b W_{\dot E \dot F \dot C b}$. Both these combinations 
have no derivatives of interest, as one sees from (\ref{ddddw}).

The other terms from (\ref{phiexp}) do not have any interesting derivatives, as
it may be easily seen from the equations in appendix \ref{appendix25}. We thus 
have shown that in $\left. \phi \right|$ there are not any derivatives of $P$ 
or $H_m$, apart from the divergence of this last field.

\subsubsection{Higher $\theta$ terms of $\phi$}
\indent

We have shown that $\left. \phi \right|$ contains no derivatives of $P$ or the 
transversal part of $H_m$. For that purpose, we needed to compute spinor 
derivatives of all the superfields of "old minimal" ${\cal N}=2$ supergravity. 
As one can see from the results of appendix \ref{appendix4}, not all spinor
derivatives of superfields originate the vector derivatives we were looking 
at; only the spinor derivatives of the "dangerous" superfields 
$\rho_{\dot A}^a$, $P$, $H_m$ and $\Phi_m^{ab}$ originate such derivatives. 

In order to compute the higher $\theta$ terms of $\phi$, we act on 
(\ref{phiexp}) with undotted spinor derivatives and follow the procedure 
described in the beginning of this section. We will find only terms that we 
have already computed but, most important, we will only  find spinor 
derivatives of the "safe" superfields $W_{AB}$, $W_{\dot A \dot B \dot C c}$, 
$X_{ab}$, $\Lambda_{\dot C c}$, $Y_{\dot A \dot B}$, $U_{A \dot A}$, 
$Y_{A B \dot C c}$, or of their spinor 
derivatives, but not of any of the "dangerous" ones. These superfields are
"safe" even in the sense that they are "closed" with respect to 
differentiation: spinor derivatives of "safe" superfields only originate 
"safe" superfields. 

This way, we conclude there are no derivatives of $P$ or the transversal 
part of $H_m$ in the ${\cal N}=2$ supersymmetrization of the fourth power
of the Weyl tensor.

\subsubsection{The behavior of $\rho_A^a$}
\indent

Our results indicate that in ${\cal N}=2$ "old minimal" supergravity with 
the ${\cal R}^4$ correction, the bosonic auxiliary fields from the compensating
nonlinear multiplet do not get derivatives and can still be eliminated. 
Auxiliary fields from the Weyl ($Y_{mn}$, $\Lambda_{Aa}$, $U_m$, $\Phi_m^{ab}$)
and vector ($X_{ab}$) multiplets get derivatives and cannot be eliminated. 

The only unclear result is the behavior of the fundamental (in terms of which
all the others are defined) auxiliary field $\rho_A^a$. Derivatives of 
$\rho_A^a$ (and of its hermitian conjugate) are not generated by the process 
of integration by parts we 
mentioned, but they are constantly being generated in the computation of
$\left. \phi \right|$ already from its beginning. This is because these 
derivatives exist already in the simple differential relations of appendix
\ref{appendix25} and in the definition (\ref{i}) of $I$. This is why we
have not fully calculated these derivatives, as we did with the other 
superfields: that would require computing a big number of terms and, for 
each term, a huge number of different contributions. This is probably
because $\rho_A^a$ belongs to a nonlinear multiplet. It would seem a miracle 
that all these derivative terms would cancel, but we have not shown that 
they do not cancel and we cannot rule it out! We can at least provide 
arguments supporting the hypothesis of cancellation. And there are at least
two good ones. The first argument is that it seems strange (although it is 
not impossible) to have a field ($\rho_A^a$) with a dynamical field equation 
while having two fields obtained from its spinorial derivatives ($P$ and $H_m$ 
- see appendix \ref{appendix24}) without such an equation. The second argument 
is that $\rho_A^a$, like $P$ and 
$H_m$, are intrinsic to the "old minimal" version of ${\cal N}=2$ supergravity;
they all come, as we saw, from the same multiplet. The physical theory 
does not depend on these auxiliary fields and, therefore, it seems natural 
that they can be eliminated from the classical theory and its quantum 
corrections. We checked that $P$ and (transversal) $H_m$ can be eliminated; the
same should be expected for $\rho_A^a$. If that was the case that the 
derivatives of $\rho_A^a$ would cancel, its field equation would be some 
function of the "dynamical" auxiliary fields and their derivatives, such that 
when replaced in their definitions in appendix \ref{appendix24}, it would 
result in differential (i.e. dynamical) field equations for these fields.

\subsection{The ${\cal N}=1$ case}
\indent

The fields of ${\cal N}=1$ conformal supergravity multiplet are the graviton 
$e_\mu^m$, the gravitino $\psi_\mu^A$ and a U(1) gauge field $A_\mu$.
(The dilation gauge field $B_\mu$ can be gauged away.) Just with these fields, 
the superconformal algebra closes off-shell. Each one of these is a 
gauge field; the corresponding gauge invariances must be considered when 
counting the number of degrees of freedom. In particular, $A_\mu$ has 4-1=3 
degrees of freedom \cite{ktvn77,pvn81}.

To obtain the "old minimal" formulation of ${\cal N}=1$ Poincar\'e 
supergravity \cite{sw78,fvn78}, we take the superconformally invariant action 
of a chiral multiplet. In order to break the superconformal and local U(1) 
invariances, one must impose some constraint which restricts the parameters of 
their transformation rules to the chiral multiplet. In superspace this is 
achieved by imposing the following nonconformal torsion constraint \cite{ht78}:
\begin{equation}
T_{Am}^{\ \ \ m}=0
\end{equation}

This constraint implies the known off-shell constraints and differential 
relations between the ${\cal N}=1$ supergravity superfields $R, G_m, W_{ABC}$:
\begin{eqnarray}
\nabla^A R&=&0 \\
\nabla^A G_{A\dot B} &=& \frac{1}{24} \nabla_{\dot B} R \label{diffg} \\
\nabla^A W_{ABC} &=& i \left( \nabla_{B \dot A} G_C^{\ \ \dot A} +\nabla_{C
\dot A} G_B^{\ \ \dot A} \right) \label{diffw}
\end{eqnarray}
which imply the relation
\begin{equation}
\nabla^2 \overline{R} - \overline{\nabla}^2 R = 96 i \nabla^n G_n 
\label{wb203}
\end{equation}

The (anti)chirality condition on $R, \overline{R}$ implies their $\theta=0$ 
components (resp. the auxiliary fields $M-iN, M+iN$) lie in antichiral/chiral 
multiplets; (\ref{diffg}) shows the spin-1/2 parts of the gravitino lie on the 
same multiplets and, according to (\ref{wb203}), so does $\partial^\mu A_\mu$
(because $\left. G_m \right|=A_m$).\footnote{The remaining scalar off-shell 
degree of freedom is the trace of the metric.}

In previous works, we have considered a similar problem to the one in the 
present paper: supersymmetrizing the fourth power of the Weyl tensor, 
${\cal W}_+^2 {\cal W}_-^2$, in the "old minimal" formulation of ${\cal N}=1$ 
supergravity 
\cite{moura01}. When we took the superspace action which included this term, we
obtained algebraic field equations for $R, \overline{R}$. According to 
(\ref{wb203}), $\nabla^n G_n$ also obbeys an algebraic equation. The auxiliary 
fields that belong to the compensating multiplet can thus still be eliminated.
This is not the case for the auxiliary fields which come from the Weyl 
multiplet ($A_m$), as we obtained, in the same work, a differential field 
equation for $G_m$. 

\subsection{Possible generalizations}
\indent

It would be interesting to figure out how the results we got can be 
generalized. Both ${\cal N}=1$ and ${\cal N}=2$ supergravities admit other 
minimal formulations, with diffe\-rent choices of compensating multiplets and 
different sets of auxiliary fields. In "new minimal" ${\cal N}=1$ \cite{sw81},
the chiral compensating multiplet is replaced by a compensa\-ting tensor 
multiplet that still breaks conformal invariance but leaves the local U(1) 
invariance unbroken. In the "new minimal" ${\cal N}=2$ \cite{wpp83}, one still
has the compensating vector multiplet that breaks conformal and local U(1)
invariances, but the nonlinear multiplet is replaced by an "improved tensor" 
compensating multiplet that breaks local SU(2) to local SO(2).

The obvious observation is that both in ${\cal N}=1$ and ${\cal N}=2$ the 
auxiliary fields from the tensor multiplets do not get derivatives in the 
supersymmetric ${\cal R}^4$ action, while the 
auxiliary fields from the Weyl (and vector in ${\cal N}=2$) multiplets do get. 
This way we notice that, in the cases we analyzed, the auxiliary fields that 
can be eliminated come from multiplets which, on-shell, have no physical 
fields; while the auxiliary fields that get derivatives come from multiplets 
with physical fiels on-shell (the graviton, the gravitino(s) and, in 
${\cal N}=2$, the vector). Our general conjecture for ${\cal R}^4$ 
supergravity, which is fully confirmed in the "old minimal" ${\cal N}=1$ case, 
can now be stated: the auxiliary fields which come from multiplets with 
on-shell physical fields cannot be eliminated, but the ones that come from 
compensating multiplets that, on shell, have no physical fields, can.

This analysis should also be extended to nonminimal versions of these theories.
These nonminimal versions would have fermionic auxiliary fields (also in 
${\cal N}=1$). This should be part of another project, and we leave more 
definite results to another work. Maybe by understanding the behavior of these 
fermionic auxiliary fields (and the ones in "new minimal" ${\cal N}=2$) we 
could fully understand what in this paper we have just conjectured: the 
behavior of the auxiliary field $\rho_A^a$ in "old minimal" ${\cal N}=2$ 
supergravity. Possibly the computations would get easier in other versions 
of the ${\cal N}=2$ theory; the fact that they were difficult in the "old
minimal" formulation, particularly the ones concerning $\rho_A^a$, is 
probably simply due to the presence of the nonlinear multiplet.

A generalization of these results to ${\cal N}=3, 4$ Poincar\'e supergravity 
theories, which can also be seen as broken superconformal theories, is more 
difficult. This is because these theories do not have an off-shell formulation 
in conventional superspace. A formulation like this could still be possible, 
but either in harmonic superspace or with multiplets with central charge.

\section{Conclusions}
\label{5}
\indent

We wrote down an action containing an ${\cal R}^4$ correction to "old minimal" 
${\cal N}=2$ supergravity. We analyzed its auxiliary field sector, and we 
concluded that the auxiliary fields belonging to the Weyl and compensating
vector multiplets acquire derivatives with these correction and cannot be
eliminated on-shell. We checked that all the terms with derivatives for the
bosonic auxiliary fields from the compensating nonlinear multiplets cancel; we
argued that the same should be true for the fermionic auxiliary field from 
this multiplet, although we have not performed the full calculation in order to
reach a definitive conclusion.

In "old minimal" ${\cal N}=1$ supergravity a similar result is valid: the 
auxiliary field from the Weyl multiplet cannot be eliminated on-shell with the
${\cal R}^4$ correction, while the ones from the chiral compensating multiplet 
can. We then conjectured that analogous results about the Weyl and compensating
multiplets should be valid for the other versions of ${\cal N}=1,2$ 
supergravity. In general, we conjecture that auxiliary fields which come from 
multiplets with on-shell physical fields cannot be eliminated, but those ones 
that come from compensating multiplets without any on shell physical fields can
be eliminated. These results should help to clarify the structure of the 
supersymmetric ${\cal R}^4$ actions in more complicated and less understood 
theories, either with more supersymmetries (in $d=4$) or in higher dimensions.

The direct supersymmetrization of higher order terms in 10 and 11 dimensions 
has been an active topic of research, although lots of questions remain open. 
Some superinvariants associated with the ${\cal R}^4$ term have been studied
\cite{nt86,dsw92,hss02,pvw00,cgnn00,gn01,nr01}, but complete supersymmetric 
effective actions including all the leading order corrections to supergravity 
are still lacking. In M-theory, because of the absence of a microscopical 
formulation, the construction of superinvariants would be even more important.
Hopefully the results we have been getting in four dimensions will provide some
insight for the higher dimensional theories!

\paragraph{Acknowledgements}
\noindent
The author is very grateful to Martin Ro\v cek for having suggested him this 
problem and, together with Warren Siegel, for very helpful discussions. He 
wishes to thank Antoine van Proeyen for correspondence and Bernard de Wit for 
discussions and for having indicated him useful references. He also wishes to 
thank INFN, Sezione di Bologna, for hospitality and financial support during a 
short visit in January 2003. This work has been partially supported 
by NSF through grant PHY-0098527.


\appendix

\section{${\cal N}=2$ SU(2) superspace conventions}
\setcounter{equation}{0}
\label{appendix1}
\indent

We work with standard SU(2) ${\cal N}=2$ superspace. We define
\begin{equation}
V_M=\left(V_m, V_{A a}, V_{\dot B b} \right)
\end{equation}
$A, \dot B$ are spinor indices, the algebra of which being exactly the same as 
the ${\cal N}=1$ case, which is fully explained in \cite{moura02,moura01}.
$a$ is an internal SU(2) index, which is raised
and lowered with an SU(2)-invariant $\varepsilon^{a b}$ tensor, just like the 
spinor indices: $T^a=\varepsilon^{a b} T_b, T_a=T^b \varepsilon_{b a}$. We take
$\varepsilon_{12}=1$. The basic rule of our conventions (different from other 
conventions in the literature) is that we use the northwest rule in every index
(spinor or SU(2)) contraction.
The complex conjugation rules are
\begin{equation}
\overline{V^a_A}=-V_{\dot A a}, \overline{V^{A a}}=V^{\dot A}_a,
\overline{V_{A a}}=V_{\dot A}^a, \overline{V_a^A}=-V^{\dot A a}, 
\overline{\varepsilon_{ab}}=\varepsilon^{a b}, \overline{\chi^{Aa} \psi^{Bb}}
= -\overline{\chi^{Aa}} \ \overline{\psi^{Bb}} \label{cc2}
\end{equation}
All other conventions regarding spacetime metrics, the Riemann tensor, Pauli 
matrices, superspace torsions and curvatures are the same as in 
\cite{moura02,moura01}.

\section{Solution to the Bianchi identities in ${\cal N}=2$ SU(2) superspace}
\label{appendix2}
\setcounter{equation}{0}
\indent

In conformal supergravity, all torsions and curvatures can be expressed in 
terms of the basic superfields $W_{AB}$, $Y_{AB}$, $U_{A \dot A}$, $X_{ab}$, 
their complex conjugates and their covariant derivatives. After breaking of 
superconformal invariance and local U(2), the basic superfields in the 
Poincar\'e theory become the physical field $W_{AB}$ and the auxiliary field 
$\rho_A^a$. All torsions and curvatures can be expressed in terms of these 
superfields, their complex conjugates and their covariant derivatives. In 
sections \ref{appendix24} and \ref{appendix25} we present the definitions of 
these superfields, and then we list the torsions and curvatures.

\subsection{Definitions}
\label{appendix24}
\indent

$\left. \rho_A^a \right|$ is an auxiliary field; $\left. W_{AB} \right|$, at 
the linearized level, is related to the field strength of the physical vector
field $A_\mu$. From (\ref{fyw}), the complete expression is 
\begin{equation}
\left. W_{AB} \right|= -\frac{i}{2 \sqrt{2}} \sigma^{mn}_{AB} F_{mn} -Y_{AB}
-\frac{i}{4} \sigma^{mn}_{AB} \left(\psi_m^{Cc} \psi_{nCc} 
+\psi_m^{\dot Cc} \psi_{n \dot Cc} \right)
\end{equation}
Now we present the definitions of the superfields of "old minimal" 
${\cal N}=2$ supergravity in terms of $W_{AB}$ and $\rho_A^a$. The hermitian 
conjugates can be easily obtained from the basic rules in (\ref{cc2}), which 
are valid for $\nabla_A^a$ and $\rho_A^a$, and the definition 
$$W_{\dot A \dot B}=\overline{W_{AB}}.$$

\begin{eqnarray}
X^{ab}&=& \frac{1}{2} \left( \nabla^{\dot A \underline{a}} -2\rho^{\dot A 
\underline{a}} \right) \rho_{\dot A}^{\underline{b}} = \frac{1}{2} 
\left( \nabla^{A \underline{a}} -2\rho^{A \underline{a}} \right) 
\rho_A^{\underline{b}} \\
Y_{AB}&=& -\frac{i}{2} \left( \nabla^a_{\underline{A}}+2
\rho^a_{\underline{A}} \right) \rho_{\underline{B} a} \\
U_{A \dot A}&=& \frac{1}{4} \left( \nabla_A^a \rho_{\dot A a} 
+\nabla_{\dot A}^a \rho_{A a} +4 \rho_A^a \rho_{\dot A a} \right) \\
\Phi_{A \dot A}^{ab}&=& \frac{i}{2} \left( \nabla_A^{\underline{a}} 
\rho_{\dot A}^{\underline{b}} - \nabla_{\dot A}^{\underline{a}} 
\rho_A^{\underline{b}} -4 \rho_A^{\underline{a}} \rho_{\dot A}^{\underline{b}}
\right) \\
P&=&i \nabla^{\dot A a} \rho_{\dot A a} \\
H_{A \dot A}&=& -i \nabla_A^a \rho_{\dot A a} +i \nabla_{\dot A}^a \rho_{A a}\\
\Lambda^{A a} &=& -i \nabla^A_b X^{ab} \\
I &=& i \nabla^{\dot A a} \Lambda_{\dot A a} - i \nabla^{A a} \Lambda_{A a} 
\label{defi} \\
W_{B C A a}&=& \frac{i}{2}\nabla_{A a} W_{B C} -\frac{i}{6} 
\left(\varepsilon_{A B} \Lambda_{C a} + \varepsilon_{A C} \Lambda_{B a} \right)
\\
Y_{B C \dot A a}&=& -\frac{i}{2} \nabla_{\dot A a}Y_{B C} \\
W_{ABCD}&=&\left(\frac{i}{4} \nabla_{\underline{A}}^b 
\nabla_{\underline{B} b} -2 Y_{\underline{AB}} \right) W_{\underline{CD}} \\
W_{\dot A \dot B \dot C \dot D}&=&\left(\frac{i}{4} 
\nabla_{\underline{\dot A}}^b \nabla_{\underline{\dot B} b} -2 
Y_{\underline{\dot A \dot B}} \right) W_{\underline{\dot C \dot D}}
\nonumber \\
P_{AB \dot C \dot D}&=&\frac{i}{8} \nabla_{\underline{A}}^b 
\nabla_{\underline{B} b} Y_{\dot C \dot D} +\frac{i}{8} 
\nabla_{\underline{\dot C}}^b \nabla_{\underline{\dot D} b} Y_{AB}
-Y_{AB} Y_{\dot C \dot D} -W_{AB} W_{\dot C \dot D} \nonumber \\
&-& U_{\underline{A} \underline{\dot C}} U_{\underline{B} \underline{\dot D}}
\\
R &=& \frac{i}{4} 
\left(\nabla^{\dot A a} \nabla^{\dot B}_a W_{\dot A \dot B}
+\nabla^{A a} \nabla^B_a W_{AB} +i \nabla^{\dot A a} \nabla_{\dot A}^b 
X_{ab} +i \nabla^{A a} \nabla_A^b X_{ab} \right) \nonumber \\
&-& 2 \left(W^{AB} Y_{AB} + W^{\dot A \dot B} Y_{\dot A \dot B} \right) - 
6 X^{ab} X_{ab} +6 U^2
\end{eqnarray}
$\left. X^{ab} \right|, \left. Y_{AB} \right|, \left. U_m \right|, 
\left. \Phi_{A \dot A}^{ab} \right|, \left. P \right|, 
\left. H_{A \dot A} \right|$ are auxiliary fields; $I$ is a dependent field. 
$W_{ABCD}$ is symmetric in all its indices, but $W_{ABCc}$ and $Y_{AB \dot Cc}$
are only symmetric in $A, B$. In the linearized approximation, 
$\left. W_{ABCc} \right|$ and $\left. Y_{AB \dot Cc} \right|$ are the gravitino
curls and $\left. W_{ABCD} \right|$, $\left. P_{AB \dot C \dot D} \right|$, 
$\left. R \right|$ are the antiself-dual Weyl tensor, the traceless Ricci 
tensor and the Ricci scalar, respectively:
\begin{eqnarray}
\left. W_{ABCc} \right|&=& -\frac{1}{4} \psi_{ABCc} +\cdots\\
\left. Y_{\dot A \dot B Cc} \right|&=& -\frac{1}{8}\psi_{\dot A \dot B Cc}
+\cdots \\
\left. W_{ABCD} \right| &=& -\frac{1}{8} {\cal W}^+_{\mu \nu \rho \sigma} 
\sigma^{\mu \nu}_{\underline{AB}} \sigma^{\rho \sigma}_{\underline{CD}} +\cdots
\label{w41} \\
\left. P_{CD \dot A \dot B} \right| &=& \frac{1}{2} 
\sigma^\mu_{\underline{C} \underline{\dot A}} 
\sigma^\nu_{\underline{D} \underline{\dot B}}
\left({\cal R}_{\mu \nu} -\frac{1}{4} g_{\mu \nu} {\cal R} \right)+\cdots \\
\left. R \right|&=& -{\cal R} +\cdots
\end{eqnarray}

\subsection{Off-shell differential relations}
\label{appendix25}
\indent

These off-shell differential relations among superfields are direct 
consequences of the Bianchi identities and the definitions in \ref{appendix24}.

\begin{eqnarray}
\nabla_A^a \rho_B^b &=& \frac{i}{4} \varepsilon_{AB} \varepsilon^{ab} 
\overline{P} -\varepsilon_{AB}
X^{ab} -i \varepsilon^{ab} Y_{AB} +2 \rho_A^b \rho_B^a \\
\nabla_A^a \rho_{\dot B}^ b&=& -\frac{i}{4} \varepsilon^{ab} H_{A \dot B} 
-\varepsilon^{ab} 
U_{A \dot B} -i \Phi_{A \dot B}^{ab} +2 \rho_A^b \rho_{\dot B}^a \\
\nabla_{\dot C c} W_{A B}&=&0, \nabla_{C c} W_{\dot A \dot B}=0 \\
\nabla_{A a} W_{B C}&=& -2i W_{B C A a} +\frac{1}{3} \left(\varepsilon_{A B} 
\Lambda_{C a} + \varepsilon_{A C} \Lambda_{B a} \right) \\
\nabla_{A a} X_{bc}&=&\frac{i}{3} \left(\varepsilon_{ab} 
\Lambda_{A c} + \varepsilon_{ac} \Lambda_{A b} \right) \\
\nabla_{A a} Y_{B C}&=&-\frac{1}{3} \left(\varepsilon_{A B} \Lambda_{C a}
+ \varepsilon_{A C} \Lambda_{B a} \right) \\
\nabla_{\dot A a} Y_{B C}&=& 2i Y_{B C \dot A a} \\
\nabla_{A a} U_{B\dot B}&=& Y_{A B \dot B a} +
\varepsilon_{A B} W^{\dot A}_{\ \ \dot B \dot A a} +\frac{2}{3}i
\varepsilon_{A B} \Lambda_{\dot B a} \\
\nabla_{\dot A}^a P &=&0, \nabla_{A}^a \overline{P} =0 \\
\nabla_{A}^a P &=& -8i W^{\ \ \ B a}_{A B} -\frac{4}{3} \Lambda_A^a 
-8 \rho^{B a} W_{A B} +2 P \rho_A^a +
2 \rho^{\dot A a} H_{A \dot A} \nonumber \\
&-& 4i \rho^{\dot A a} U_{A \dot A} -8 \nabla_{A \dot A} \rho^{\dot A a}
-8 \Phi_{A \dot A}^{ab} \rho^{\dot A}_b \\
\nabla_{A}^a H_{B \dot B}&=& 8i \varepsilon_{AB} 
W_{\dot A \dot B}^{\ \ \dot A a} +\frac{4}{3} \varepsilon_{AB}
\Lambda_{\dot B}^a -4 \varepsilon_{AB} \rho^{\dot A a} W_{\dot A \dot B}
-4i \varepsilon_{AB} \rho_{\dot B b} X^{ba} \nonumber \\
&-& 2 \varepsilon_{AB} \overline{P} \rho_{\dot B}^a
-2 \varepsilon_{AB} \rho^{C a} H_{C \dot B} +4 \rho_{\dot B}^a Y_{AB}
+2i \rho_A^a U_{B \dot B} \nonumber \\ 
&-& 8 \rho_{A b} \Phi_{B \dot B}^{ab} + 8 \rho_{B b} \Phi_{A \dot B}^{ab}
-4 \nabla_{B \dot B} \rho^a_A +8 \nabla_{A \dot B} \rho^a_B \\
\nabla_{A}^a \Phi_{B \dot B}^{bc} &=& 2i \varepsilon_{AB} 
\varepsilon^{a \underline{b}} W_{\dot B \dot A}^{\ \ \dot A \underline{c}}
+ \frac{2}{3} \varepsilon_{AB} \varepsilon^{a \underline{b}}
\Lambda^{\underline{c}}_{\dot B} -2 \varepsilon_{AB} 
\varepsilon^{a \underline{b}} \rho^{\dot A \underline{c}} W_{\dot A \dot B} 
\nonumber \\
&+& 2i \varepsilon^{a \underline{b}} Y_{AB \dot B}^{\ \ \ \ \underline{c}} 
-2 \varepsilon^{a \underline{b}} Y_{AB} \rho^{\underline{c}}_{\dot B} 
+i \varepsilon^{a \underline{b}} \rho^{\underline{c}}_A U_{B \dot B}
-2i \varepsilon^{a \underline{b}} \rho^{\underline{c}}_B U_{A \dot B}
\nonumber \\
&-& 4 \varepsilon^{a \underline{b}} \rho_{A d} \Phi_{B \dot B}^{d\underline{c}}
-2 \varepsilon^{a \underline{b}} \nabla_{B \dot B} \rho^{\underline{c}}_A
-2i \varepsilon_{AB} X^{a \underline{b}} \rho^{\underline{c}}_{\dot B}
+2 \rho^{\underline{b}}_A \Phi_{B \dot B}^{a \underline{c}} \\
\nabla_A^a \Lambda_B^b &=& \frac{i}{8} I \varepsilon_{AB} \varepsilon^{ab} 
+ \frac{3}{2} \varepsilon_{AB} \varepsilon^{ab} \nabla_{C \dot C} U^{C \dot C}
\nonumber \\
&+& \frac{3}{2}i \varepsilon_{AB} \varepsilon^{ab} \left( Y^{CD} W_{CD} -
Y^{\dot C \dot D} W_{\dot C \dot D} \right) -6 Y_{AB} X^{ab} \\
\nabla_A^a \Lambda_{\dot B}^b &=& 3i \varepsilon^{ab} \left( \nabla_{B \dot B}
W_A^{\ B} + \nabla_A^{\ \dot A} Y_{\dot A \dot B} \right) -3 \nabla_{A \dot B}
X^{ab} -6i U_{A \dot B} X^{ab} \\
\nabla_{\dot B b} W_{\dot C \dot D \dot A a} &=& \varepsilon_{ba} 
W_{\dot C \dot D \dot A \dot B} +\frac{1}{12} \varepsilon_{ba}
\left( \varepsilon_{\dot A \dot C} \varepsilon_{\dot B \dot D} + 
\varepsilon_{\dot A \dot D} \varepsilon_{\dot B \dot C} \right) R
- \varepsilon_{\dot B \dot A}
\nabla^C_{\ \underline{\dot C}} \Phi_{Cb \underline{\dot D}a} \nonumber \\
&+& i \varepsilon_{ba} 
\varepsilon_{\underline{\dot C} \underline{\underline{\dot A}}}
\nabla^C_{\ \underline{\dot D}}  U_{C \underline{\underline{\dot B}}}
+ 2 \varepsilon_{\dot B \dot A} W_{\dot C \dot D \dot E \underline{b}} 
\rho^{\dot E}_{\underline{a}} - 2 \varepsilon_{\dot B \dot A}
Y_{\dot C \dot D E \underline{b}} \rho^E_{\underline{a}} \nonumber \\
&+&\varepsilon_{\dot B \dot A} \Phi^{Ce}_{\ \ \underline{\dot C} \underline{b}}
\Phi_{Ce \underline{\dot D} \underline{a}} +2 \varepsilon_{ba} 
Y_{\underline{\dot C} \dot B} W_{\underline{\dot D} \dot A} +\varepsilon_{ba}
\varepsilon_{\underline{\dot C} \dot B} \varepsilon_{\underline{\dot D} \dot A}
W^{CD} Y_{CD} \nonumber \\
&+&\varepsilon_{ba}
\varepsilon_{\underline{\dot C} \dot B} \varepsilon_{\underline{\dot D} \dot A}
X^2 +\varepsilon_{ba}
\varepsilon_{\underline{\dot C} \dot B} \varepsilon_{\underline{\dot D} \dot A}
U^2 +2i \varepsilon_{\underline{\dot D} \dot A} Y_{\underline{\dot C} \dot B}
X_{ba} \nonumber \\
&+& 2i \varepsilon_{\underline{\dot C} \dot B} W_{\underline{\dot D} \dot A}
X_{ba} \\
\nabla_{Bb} W_{\dot C \dot D \dot A a} &=& \varepsilon_{ab} \nabla_{B \dot A}
W_{\dot C \dot D} + \varepsilon_{ab} \varepsilon_{\dot A \underline{\dot C}} 
\nabla_B^{\ \dot B} W_{\dot B \underline{\dot D}} - \varepsilon_{ab} 
\varepsilon_{\dot A \underline{\dot C}} \nabla^C_{\ \underline{\dot D}} Y_{BC} 
\nonumber \\
&+&i \varepsilon_{\dot A \underline{\dot C}} \nabla_{B \underline{\dot D}} 
X_{ab} - 2 \varepsilon_{\dot A \underline{\dot C}} U_{B \underline{\dot D}} 
X_{ab} + i \varepsilon_{ab} U_{B \dot A} W_{\dot C \dot D} \nonumber \\
&+& 2i \varepsilon_{ab} \varepsilon_{\dot A \underline{\dot C}} U_B^{\ \dot B} 
W_{\dot B \underline{\dot D}} \\
\nabla_{\dot B b} Y_{C D \dot A a} &=& \varepsilon_{ba} P_{C D \dot B \dot A}
+i \varepsilon_{ba} \nabla_{\underline{C} \underline{\dot A}} 
U_{\underline{D} \underline{\dot B}} - \varepsilon_{\dot B \dot A}
\nabla^{\ \ \dot C}_{ \underline{C}} \Phi_{\underline{D} a \dot C b}
\nonumber \\
&-& 2 \varepsilon_{\dot B \dot A} W_{C D E \underline{b}} 
\rho^E_{\underline{a}} + 2 \varepsilon_{\dot B \dot A}
Y_{C D \dot E \underline{b}} \rho^{\dot E}_{\underline{a}} 
+\varepsilon_{\dot B \dot A} \Phi^{\ \ \ \dot C e}_{ \underline{C} 
\underline{b}} \Phi_{\underline{D} \underline{a} \dot C e} \nonumber \\
&+& \varepsilon_{ba} W_{CD} W_{\dot B \dot A} + \varepsilon_{ba} Y_{CD} 
Y_{\dot B \dot A} -\varepsilon_{ba} \varepsilon_{\dot B \dot A}
W_{\underline{C}}^{\ E} Y_{E \underline{D}} \nonumber \\
&-& i \varepsilon_{\dot B \dot A} X_{ab} Y_{CD} + i \varepsilon_{\dot B \dot A}
X_{ab} W_{CD} +\varepsilon_{ba} U_{\underline{C} \underline{\dot A}} 
U_{\underline{D} \underline{\dot B}} \label{dbardbary}\\
\nabla_{B b} Y_{C D \dot A a} &=& \varepsilon_{ba} \varepsilon_{B\underline{C}}
\nabla_{\underline{D}}^{\ \ \dot C} W_{\dot C \dot A} - \varepsilon_{ba}
\nabla_{\underline{D} \dot A} Y_{B\underline{C}} +i\varepsilon_{B\underline{C}}
\nabla_{\underline{D} \dot A} X_{ab} \nonumber \\
&-& i \varepsilon_{ba} Y_{CD} U_{B \dot A} +2i \varepsilon_{ba} 
Y_{B\underline{C}} U_{\underline{D} \dot A} +2 \varepsilon_{B\underline{C}} 
U_{\underline{D} \dot A} X_{ba} \label{ddbary} \\
\nabla_{\dot E e} W_{\dot A \dot B \dot C \dot D} &=&-2i
\varepsilon_{\dot E \underline{\dot C}} \nabla^E_{\ \underline{C}} 
Y_{\underline{\dot A} \underline{\dot B} E e} -4i
Y_{\dot E \underline{\dot C}}
W_{\underline{\dot A} \underline{\dot B} \underline{\dot D} e}
-8 \varepsilon_{\dot E \underline{\dot C}} X_e^{\ f} 
W_{\underline{\dot A} \underline{\dot B} \underline{\dot D} f} \nonumber \\
&+& 3 \varepsilon_{\dot E \underline{\dot C}} 
Y_{\underline{\dot A} \underline{\dot B} E e} U^E_{\ \underline{\dot D}}
-2i W_{\underline{\dot A} \underline{\dot B} \dot E e}
Y_{\underline{\dot C} \underline{\dot D}} \\
\nabla_{A a} W_{\dot A \dot B \dot C \dot D} &=& 2i\nabla_{A\underline{\dot C}}
W_{\underline{\dot A} \underline{\dot B} \underline{\dot D} a} +
U_{A\underline{\dot C}} W_{\underline{\dot A} \underline{\dot B} 
\underline{\dot D} a} +4i W_{\underline{\dot C} \underline{\dot D}}
Y_{\underline{\dot A} \underline{\dot B} A a} \\
\nabla_{\dot E e} P_{C D \dot B \dot A} &=& 2i \nabla_{\underline{C} \dot E}
Y_{\dot A \dot B \underline{D} e} -i \nabla_{\underline{C} \underline{\dot A}}
Y_{\dot E \underline{\dot B} \underline{D} e} +i 
\varepsilon_{\dot E\underline{\dot B}} \nabla_{\underline{C}\underline{\dot A}}
W_{\underline{D} E \ e}^{\ \ \ E}- 2iY_{CD \dot E e}Y_{\dot A \dot B} 
\nonumber \\
&-&i Y_{\dot E\underline{\dot A}} Y_{CD \underline{\dot B} e}
+i\varepsilon_{\dot E\underline{\dot A}}W_{\underline{C}}^{\ E} 
Y_{E \underline{D} \underline{\dot B} e}- 2i W_{CD} W_{\dot A \dot B \dot E e} 
\nonumber \\
&-&i\varepsilon_{\dot E\underline{\dot A}} W_{CD} 
W_{\underline{\dot B} \dot C \ e}^{\ \ \ \dot C}
-5 \varepsilon_{\dot E\underline{\dot A}} X_e^{\ f} Y_{CD \underline{\dot B} f}
-2\varepsilon_{\dot E\underline{\dot A}} W_{CDF e} U^F_{\ \underline{\dot B}}
\nonumber \\
&+& \varepsilon_{\dot E\underline{\dot A}} U_{\underline{C} \underline{\dot B}}
W_{\underline{D} E \ e}^{\ \ \ E} +\frac{1}{2} U_{\underline{C} 
\underline{\dot B}} Y_{\dot E \underline{\dot A} \underline{D} e}
+ 2 U_{\underline{C} \dot E} Y_{\dot A \dot B \underline{D} e} \\
\nabla_{\dot A a} R &=& -i \nabla^{B \dot B} Y_{\dot A \dot B B a}
+3i \nabla^B_{\ \dot A} W_{B C \ a}^{\ \ \ C} +10 X_{ab} 
W_{\dot A \dot B}^{\ \ \ \dot B b} \nonumber
- 2i W_{\dot C \dot B \ a}^{\ \ \ \dot B} Y_{\dot A}^{\ \dot C} \\
&-& 4i W_{\dot C \dot B \dot A a}Y^{\dot C \dot B}-3iW^{CB}Y_{CB \dot A a}
+ \frac{9}{2} U^{B \dot B} Y_{\dot A \dot B B a}
+ \frac{9}{2} U^B_{\ \dot A} W_{B C a}^{\ \ \ C} \label{dbarr}
\end{eqnarray}
Using (\ref{dbardbary}) and (\ref{ddbary}), one may compute 
$\nabla^{B \dot B} Y_{\dot A \dot B B a}$; replacing in (\ref{dbarr}), we get 
the more convenient expression
\begin{eqnarray}
\nabla_{\dot A a} R &=& 4i \nabla^B_{\ \dot A} W_{B C \ a}^{\ \ \ C} 
+12 X_{ab} W_{\dot A \dot B}^{\ \ \ \dot B b} -2iW^{CB}Y_{CB \dot A a}
\nonumber \\
&-& 6iW_{\dot C \dot B \dot A a}Y^{\dot C \dot B}+ 12 U^{B \dot B} 
Y_{\dot A \dot B B a} + 4 U^B_{\ \dot A} W_{B C \ a}^{\ \ \ C}
\end{eqnarray}

\subsection{Torsions}

\begin{eqnarray}
T_{A\dot B}^{a b m}&=&-2i\varepsilon^{a b} \sigma_{A \dot B}^m \nonumber \\
T_{A B}^{a b m}, T_{\dot A \dot B}^{a b m}&=& 0 \nonumber \\
T_{\dot A a B b C c}, T_{A a \dot B b \dot C c} &=& 0 \nonumber \\
T_{A a B b C c}, T_{\dot A a \dot B b \dot C c} &=& 0 \nonumber \\
T_{A}^{a m n}, T_{\dot A}^{a m n} &=&0 \nonumber \\
T_{A \dot A B b C c} &=& -\frac{i}{2} \varepsilon_{b c}
\left( \varepsilon_{A B} U_{C \dot A} + \varepsilon_{A C} U_{B \dot A}
\right) \nonumber \\
T_{A \dot A \dot B b \dot C c} &=& \frac{i}{2} \varepsilon_{b c}
\left( \varepsilon_{\dot A \dot B} U_{A \dot C} +
\varepsilon_{\dot A \dot C} U_{A \dot B} \right) \nonumber \\
T_{A \dot A B b \dot C c} &=& -\varepsilon_{b c} \left( \varepsilon_{A B} 
W_{\dot A \dot C} + \varepsilon_{\dot A \dot C} Y_{A B} \right) -i
\varepsilon_{A B} \varepsilon_{\dot A \dot C} X_{b c} \nonumber \\
T_{A \dot A \dot B b C c} &=& \varepsilon_{b c} \left(\varepsilon_{A C}
Y_{\dot A \dot B} + \varepsilon_{\dot A \dot B} W_{A C} \right) 
+i \varepsilon_{A C} \varepsilon_{\dot A \dot B} X_{b c} \nonumber \\
T_{m n p}&=&0 \nonumber \\
T_{A \dot A B \dot B C c} &=& -\varepsilon_{\dot A \dot B} W_{A B Cc}
- \varepsilon_{A B} Y_{\dot A \dot B Cc}
\nonumber \\
T_{A \dot A B \dot B \dot C c} &=& \varepsilon_{A B}
W_{\dot A \dot B \dot Cc} +\varepsilon_{\dot A \dot B} Y_{A B \dot Cc}
\end{eqnarray}

\subsection{Lorentz curvatures}

\begin{eqnarray}
R_{Aa Bb CD}&=& -2i \varepsilon_{AB} \varepsilon_{ab} Y_{CD}
+2 \left(\varepsilon_{AC} \varepsilon_{BD} +\varepsilon_{AD} 
\varepsilon_{BC} \right) X_{ab} 
\nonumber \\
R_{Aa Bb \dot C \dot D} &=& -2i \varepsilon_{AB} \varepsilon_{ab} 
W_{\dot C \dot D} 
\nonumber \\
R_{\dot A a \dot B b CD} &=& -2i \varepsilon_{\dot A \dot B}
\varepsilon_{ab} W_{CD} 
\nonumber \\
R_{\dot A a \dot B b \dot C \dot D} &=& -2i \varepsilon_{\dot A \dot B} 
\varepsilon_{ab} Y_{\dot C \dot D} +2 \left(\varepsilon_{\dot A \dot C} 
\varepsilon_{\dot B \dot D} +\varepsilon_{\dot A \dot D} 
\varepsilon_{\dot B \dot C} \right) X_{ab} 
\nonumber \\
R_{\dot Aa Bb CD} &=& \varepsilon_{ab} \left(\varepsilon_{BC} U_{D \dot A}
+ \varepsilon_{BD} U_{C \dot A} \right) 
\nonumber \\
R_{\dot Aa Bb \dot C \dot D} &=& -\varepsilon_{ab} 
\left(\varepsilon_{\dot A \dot C} U_{B \dot D} + \varepsilon_{\dot A \dot D}
U_{B \dot C} \right) 
\nonumber \\
R_{A \dot A B b CD}&=&-i \varepsilon_{B\underline{C}} 
Y_{A\underline{D} \dot A b} +i \varepsilon_{AB} 
Y_{CD \dot A b} +\frac{i}{2} \left(\varepsilon_{AC} \varepsilon_{BD} 
+\varepsilon_{AD} \varepsilon_{BC} \right) W_{\dot A \dot B \ b}^{\ \ \ \dot B}
\nonumber \\
R_{A \dot A \dot B b CD}&=& i \varepsilon_{A \underline{C}} 
Y_{\dot A \dot B \underline{D} b} +2i \varepsilon_{\dot A \dot B} 
W_{\underline{ACD} b} +\frac{i}{3} \varepsilon_{\dot A \dot B}
\varepsilon_{A \underline{C}} W_{\underline{D} \ B b}^{\ B} \nonumber \\
R_{A \dot A B b \dot C \dot D} &=& i \varepsilon_{\dot A \underline{\dot C}} 
Y_{AB \underline{\dot D} b} +2i \varepsilon_{AB} 
W_{\underline{\dot A \dot C \dot D} b} +\frac{i}{3} \varepsilon_{AB}
\varepsilon_{\dot A \underline{\dot C}} 
W_{\underline{\dot D} \ \dot B b}^{\ \dot B} \nonumber \\
R_{A \dot A \dot B b \dot C \dot D} &=&-i 
\varepsilon_{\dot B \underline{\dot C}} 
Y_{\dot A \underline{\dot D} A b} +i \varepsilon_{\dot A \dot B} 
Y_{\dot C \dot D A b} +\frac{i}{2} \left(\varepsilon_{\dot A \dot C} 
\varepsilon_{\dot B \dot D} + \varepsilon_{\dot A \dot D} 
\varepsilon_{\dot B \dot C} \right) W_{AB\ b}^{\ \ \ B}
\nonumber \\
R_{A \dot A B \dot B CD}&=& \varepsilon_{AB} P_{CD \dot A \dot B} +
\frac{1}{12} \varepsilon_{\dot A \dot B} \left(\varepsilon_{AC} 
\varepsilon_{BD} + \varepsilon_{AD} \varepsilon_{BC} \right) R
+\varepsilon_{\dot A \dot B} W_{ABCD} 
\nonumber \\
R_{A \dot A B \dot B \dot C \dot D} &=& \varepsilon_{\dot A \dot B} 
P_{AB \dot C \dot D} + \frac{1}{12} \varepsilon_{AB} \left(
\varepsilon_{\dot A \dot C} \varepsilon_{\dot B \dot D} + 
\varepsilon_{\dot A \dot D} \varepsilon_{\dot B \dot C} \right) R
+\varepsilon_{AB} W_{\dot A \dot B \dot C \dot D}
\end{eqnarray}

\subsection{SU(2) curvatures}

\begin{eqnarray}
R_{Aa Bb cd}&=& -2 \varepsilon_{AB} \left(\varepsilon_{bd} X_{ac}-
\varepsilon_{ad} X_{bc} \right) -2i \left(\varepsilon_{ac} \varepsilon_{bd} 
+ \varepsilon_{ad} \varepsilon_{bc} \right) Y_{AB} 
\nonumber \\
R_{\dot A a \dot B bcd}&=& -2 \varepsilon_{\dot A \dot B} 
\left(\varepsilon_{bd} X_{ac}- \varepsilon_{ad} X_{bc} \right) 
-2i \left(\varepsilon_{ac} \varepsilon_{bd} + \varepsilon_{ad} 
\varepsilon_{bc} \right) Y_{\dot A \dot B} 
\nonumber \\
R_{A a\dot B b cd}&=&-2 \left(\varepsilon_{ac} \varepsilon_{bd} 
+ \varepsilon_{ad} \varepsilon_{bc} \right) U_{A \dot B} 
\nonumber \\
R_{A \dot A Bb cd} &=&2i \varepsilon_{b\underline{d}} 
Y_{AB\dot A \underline{c}} -2i \varepsilon_{AB} \varepsilon_{b\underline{d}}
W_{\dot A \dot B \ \underline{c}}^{\ \ \ \dot B} - \frac{2}{3} \varepsilon_{AB}
\varepsilon_{b\underline{d}} \Lambda_{\dot A \underline{c}} \nonumber \\
R_{A \dot A \dot B b cd} &=& 2i \varepsilon_{b\underline{d}}
Y_{\dot A \dot B A \underline{c}} -2i \varepsilon_{\dot A \dot B}
\varepsilon_{b\underline{d}} W_{AB \ \underline{c}}^{\ \ \ B}
- \frac{2}{3} \varepsilon_{\dot A \dot B}
\varepsilon_{b\underline{d}} \Lambda_{A \underline{c}} \nonumber \\
R_{A \dot A B \dot B cd} &=& -2 \varepsilon_{\dot A \dot B}
\left(W_{ABE \underline{c}} \rho^E_{\underline{d}} -
Y_{AB\dot E \underline{c}} \rho^{\dot E}_{\underline{d}} +
\frac{1}{2} \nabla_{\underline{A}}^{\ \dot E} \Phi_{\underline{B} \dot E cd}
-\frac{1}{2} \Phi_{\underline{A} \ \ \underline{c}}^{\ \dot E e}
\Phi_{\underline{B} \dot E \underline{d}e} \right) \nonumber \\
&+& 2 \varepsilon_{AB} \left(
W_{\dot A \dot B \dot E \underline{c}} \rho^{\dot E}_{\underline{d}} 
- Y_{\dot A \dot B E \underline{c}} \rho^E_{\underline{d}} -\frac{1}{2}
\nabla^E_{\ \underline{\dot A}} \Phi_{E \underline{\dot B} cd} +
\frac{1}{2} \Phi_{\ \underline{\dot A} \underline{c}e}^E
\Phi_{E \underline{\dot B} \underline{d}}^{\ \ \ \ e}
\right) \nonumber
\end{eqnarray}

We see that all torsions and curvatures can be expressed in 
terms of the basic superfields $W_{AB}$, $Y_{AB}$, 
$U_{A \dot A}$, $X_{ab}$, their complex conjugates and their covariant 
derivatives (see section \ref{appendix24}). This is obvious except for 
$R_{A \dot A B \dot B cd}$, which may be rewritten as 
\begin{eqnarray}
R_{A \dot A B \dot B cd} &=& \varepsilon_{\dot A \dot B} \left( \frac{1}{2}
\nabla^{\dot C}_{\ \underline{c}} Y_{AB \dot C \underline{d}} +i X_{cd} Y_{AB}
-i X_{cd} W_{AB} \right) \nonumber \\
&+& \varepsilon_{AB} \left(\frac{1}{2} \nabla^C_{\ \underline{c}} 
Y_{\dot A \dot B C \underline{d}} +i X_{cd} Y_{\dot A \dot B} -i X_{cd} 
W_{\dot A \dot B} \right)
\end{eqnarray}


\section{The lagrangian in components}
\label{appendix3}
\setcounter{equation}{0}
\indent


\subsection{Calculation of $\nabla_{\dot Cc} \nabla_{\dot Bb} 
\nabla_{\dot Aa} \overline{W}^2$}
\label{appendix31}
\indent

In this section, we fully express the terms with two spinorial derivatives 
arising in the calculation of (\ref{d3w2}) as functions of those with one 
spinorial derivative, previously computed in appendix \ref{appendix2}. These
terms are:
\begin{eqnarray}
\nabla_{\dot Cc} \nabla_{\dot Bb} W_{\dot A \dot D} &=& 2i \nabla_{\dot Cc} 
W_{\dot A \dot D \dot Bb} - \frac{2}{3} \varepsilon_{\dot B \underline{\dot A}}
\nabla_{\dot Cc} \Lambda_{\underline{\dot D} b} \\
\nabla_{\dot Cc} \nabla_{\dot Bb} W_{\dot D \dot E \dot A a} &=&-
\varepsilon_{a b} \nabla_{\dot Cc} W_{\dot D \dot E \dot A \dot B} - 2
\varepsilon_{a b} \left(\nabla_{\dot Cc} Y_{\dot B \underline{\dot D}} \right)
W_{\underline{\dot E} \dot A} -2 \varepsilon_{a b} 
Y_{\dot B \underline{\dot D}} \nabla_{\dot Cc} W_{\underline{\dot E} \dot A}
\nonumber \\
&+& 2 \varepsilon_{\dot B \dot A} \left(\nabla_{\dot Cc} 
W_{\dot D \dot E \dot F \underline{a}} \right) \rho_{\underline{b}}^{\dot F}
-2 \varepsilon_{\dot B \dot A} W_{\dot D \dot E \dot F \underline{a}}
\nabla_{\dot Cc} \rho_{\underline{b}}^{\dot F} \nonumber \\
&-& 2 \varepsilon_{\dot B \dot A} \left(\nabla_{\dot Cc} 
Y_{\dot D \dot E E \underline{a}} \right) \rho_{\underline{b}}^E 
+2 \varepsilon_{\dot B \dot A} Y_{\dot D \dot E E \underline{a}}
\nabla_{\dot Cc} \rho_{\underline{b}}^E - \varepsilon_{\dot B \dot A} 
\nabla_{\dot Cc} \nabla^E_{\ \underline{\dot D}} \Phi_{E \underline{\dot E} ab}
\nonumber \\
&+&2 \varepsilon_{\dot B \dot A} 
\Phi_{\ \underline{\dot D} \ \ \underline{a}}^{E \ \ e}
\nabla_{\dot Cc} \Phi_{E \underline{\dot E} e \underline{b}}
-2i \varepsilon_{\dot B \underline{\dot D}} \left(\nabla_{\dot Cc} X_{ab} 
\right) W_{\underline{\dot E} \dot A} \nonumber \\
&-&2i \varepsilon_{\dot B \underline{\dot D}} X_{ab} \nabla_{\dot Cc} 
W_{\underline{\dot E} \dot A}+\frac{1}{6} \varepsilon_{ab} 
\varepsilon_{\underline{\dot E} \dot A} \varepsilon_{\dot B \underline{\dot D}}
\nabla_{\dot Cc} R \nonumber \\
&-&i \varepsilon_{ab} 
\varepsilon_{\underline{\dot E} \underline{\underline{\dot A}}}
\nabla_{\dot Cc} \nabla^E_{\ \underline{\dot D}} 
U_{E \underline{\underline{\dot B}}} +\varepsilon_{ab} 
\varepsilon_{\underline{\dot E} \dot A} \varepsilon_{\dot B \underline{\dot D}}
U^{F \dot F} \nabla_{\dot Cc} U_{F \dot F} \nonumber \\
&+&2i \varepsilon_{\underline{\dot E} \dot A} \left(\nabla_{\dot Cc} 
Y_{\dot B \underline{\dot D}} \right) X_{ab} 
+2i \varepsilon_{\underline{\dot E} \dot A} Y_{\dot B \underline{\dot D}}
\nabla_{\dot Cc} X_{ab}  \nonumber \\
&+&\varepsilon_{ab} \varepsilon_{\underline{\dot E} \dot A} 
\varepsilon_{\dot B \underline{\dot D}} W^{AB} \nabla_{\dot Cc} Y_{AB}
+2\varepsilon_{ab} \varepsilon_{\underline{\dot E} \dot A} 
\varepsilon_{\dot B \underline{\dot D}} X^{de} \nabla_{\dot Cc} X_{de} \\
\nabla_{\dot Cc} \nabla_{\dot Bb} \Lambda_{\dot Aa} &=& \frac{i}{8}
\varepsilon_{ab} \varepsilon_{\dot B \dot A} \nabla_{\dot Cc} I -\frac{3}{2}
\varepsilon_{ab} \varepsilon_{\dot B \dot A} \nabla_{\dot Cc} \nabla^{F \dot F}
U_{F \dot F} -\frac{3}{2}i \varepsilon_{ab} \varepsilon_{\dot B \dot A} 
W^{AB} \nabla_{\dot Cc} Y_{AB} \nonumber \\ 
&+&\frac{3}{2}i \varepsilon_{ab} \varepsilon_{\dot B \dot A} W^{\dot E \dot F} 
\nabla_{\dot Cc} Y_{\dot E \dot F} +\frac{3}{2}i \varepsilon_{ab} 
\varepsilon_{\dot B \dot A} Y^{\dot E \dot F} \nabla_{\dot Cc} 
W_{\dot E \dot F} +6 X_{ab} \nabla_{\dot Cc} Y_{\dot B \dot A} \nonumber \\
&+&6 Y_{\dot B \dot A} \nabla_{\dot Cc} X_{ab} \label{ddlambda}
\end{eqnarray}
For (\ref{ddlambda}) we need $\nabla_{\dot Cc} I$, which we compute here.

\begin{eqnarray}
\nabla_{\dot Cc} I &=& 4\nabla_{\dot Cc} R
-48 X^{ab} \nabla_{\dot Cc} X_{ab} -12 W^{AB} \nabla_{\dot Cc} Y_{AB} -12 
W^{\dot A \dot B} \nabla_{\dot Cc} Y_{\dot A \dot B} \nonumber \\
&-& 12 Y^{\dot A \dot B} \nabla_{\dot Cc} W_{\dot A \dot B} -24 U^{F \dot F} 
\nabla_{\dot Cc} U_{F \dot F} +3 P \nabla_{\dot Cc} \overline{P}
+ 3 H^{F \dot F} \nabla_{\dot Cc} H_{F \dot F} \nonumber \\
&-&24 \Phi^{F \dot F}_{ab} \nabla_{\dot Cc} \Phi^{ab}_{F \dot F}
-6 \nabla_{\dot Cc} \nabla^{F \dot F} H_{F \dot F} -16i \left(\nabla_{\dot Cc} 
\rho^{Aa} \right) \Lambda_{Aa} + 16i \rho^{Aa} \nabla_{\dot Cc} \Lambda_{Aa} 
\nonumber \\
&+&16i \left(\nabla_{\dot Cc} \rho^{\dot Aa} \right) \Lambda_{\dot Aa}
-16i \rho^{\dot Aa} \nabla_{\dot Cc} \Lambda_{\dot Aa} +48 
\left(\nabla_{\dot Cc} \rho^{Aa} \right) W_{AB\ a}^{\ \ \ B} \nonumber \\
&-&48 \rho^{Aa} \nabla_{\dot Cc} W_{AB\ a}^{\ \ \ B} -48\left(\nabla_{\dot Cc} 
\rho^{\dot Aa} \right) W_{\dot A \dot B\ a}^{\ \ \ \dot B} +48 \rho^{\dot Aa} 
\nabla_{\dot Cc} W_{\dot A \dot B\ a}^{\ \ \ \dot B} \nonumber \\
&-&96i W^{AB} \rho_{Aa} \nabla_{\dot Cc} \rho^a_B -96i W^{\dot A \dot B} 
\rho_{\dot Aa} \nabla_{\dot Cc} \rho^a_{\dot B} +48i \rho_{\dot Aa} 
\rho^a_{\dot B} \nabla_{\dot Cc} W^{\dot A \dot B} \nonumber \\
&-&48 U^{A \dot A} \left(\nabla_{\dot Cc} \rho_{Aa} \right) \rho^a_{\dot A}
+48 U^{A \dot A} \rho_{Aa} \nabla_{\dot Cc} \rho^a_{\dot A}
-48 \rho_{Aa} \rho^a_{\dot A} \nabla_{\dot Cc} U^{A \dot A} \nonumber \\
&+&48i \left(\nabla_{\dot Cc} \rho_{Aa} \right) \nabla^{A \dot A} 
\rho^a_{\dot A}- 48i \rho_{Aa} \nabla_{\dot Cc} \nabla^{A \dot A} 
\rho^a_{\dot A} -48i \left(\nabla_{\dot Cc} \rho_{\dot Aa} \right) 
\nabla^{A \dot A} \rho^a_A \nonumber \\
&+&48i \rho_{\dot Aa} \nabla_{\dot Cc} \nabla^{A \dot A} \rho^a_A
+96i \Phi^{ab}_{A \dot A} \rho^{\dot A}_b \nabla_{\dot Cc} \rho^A_a
+96i \Phi^{ab}_{A \dot A} \rho^A_b \nabla_{\dot Cc} \rho^{\dot A}_a\nonumber \\
&+&96i \rho^A_a \rho^{\dot A}_b \nabla_{\dot Cc} \Phi^{ab}_{A \dot A} 
\label{di}
\end{eqnarray}
Terms involving combinations of vector and spinor covariant derivatives may be 
written as vector derivatives of the relations listed in appendix 
\ref{appendix2} using the commutation relations.

\subsection{Calculation of $\nabla_{\dot Dd} \nabla_{\dot Cc} 
\nabla_{\dot Bb} \nabla_{\dot Aa} \overline{W}^2$}
\label{appendix32}
\indent

In this section, we fully express the terms with two and three spinorial 
derivatives arising in the calculation of (\ref{d4w2}) as functions of those 
with one spinorial derivative, previously computed in appendix \ref{appendix2}.
These terms are:
\begin{eqnarray}
\nabla_{\dot Dd} \nabla_{\dot Cc} \nabla_{\dot Bb} W_{\dot E \dot F} &=& 
2i\nabla_{\dot Dd} \nabla_{\dot Cc} W_{\dot E \dot F \dot Bb} - \frac{2}{3} 
\varepsilon_{\dot B \underline{\dot E}} \nabla_{\dot Dd} \nabla_{\dot Cc} 
\Lambda_{\underline{\dot F} b} \\
\nabla_{\dot Dd} \nabla_{\dot Cc} \nabla_{\dot Bb} W_{\dot E \dot F \dot A a} 
&=&- \varepsilon_{a b}\nabla_{\dot Dd}  \nabla_{\dot Cc} 
W_{\dot E \dot F \dot A \dot B} - 2 \varepsilon_{a b} \left(\nabla_{\dot Dd} 
\nabla_{\dot Cc} Y_{\dot B \underline{\dot E}} \right) 
W_{\underline{\dot F} \dot A} \nonumber \\
&+&2 \varepsilon_{a b} \left(\nabla_{\dot Cc} Y_{\dot B \underline{\dot E}} 
\right) \nabla_{\dot Dd} W_{\underline{\dot F} \dot A}
-2 \varepsilon_{a b} \left(\nabla_{\dot Dd} Y_{\dot B \underline{\dot E}} 
\right) \nabla_{\dot Cc} W_{\underline{\dot F} \dot A} \nonumber \\
&-&2 \varepsilon_{a b} Y_{\dot B \underline{\dot E}} \nabla_{\dot Dd} 
\nabla_{\dot Cc} W_{\underline{\dot F} \dot A} +2 \varepsilon_{\dot B \dot A} 
\left(\nabla_{\dot Dd} \nabla_{\dot Cc} 
W_{\dot E \dot F \dot G \underline{a}} \right) \rho_{\underline{b}}^{\dot G}
\nonumber \\
&+&2 \varepsilon_{\dot B \dot A} \left(\nabla_{\dot Cc} 
W_{\dot E \dot F \dot G \underline{a}} \right) \nabla_{\dot Dd} 
\rho_{\underline{b}}^{\dot G} -2 \varepsilon_{\dot B \dot A} 
\left(\nabla_{\dot Dd} W_{\dot E \dot F \dot G \underline{a}} \right) 
\nabla_{\dot Cc} \rho_{\underline{b}}^{\dot G} \nonumber \\
&+&2 \varepsilon_{\dot B \dot A} W_{\dot E \dot F \dot G \underline{a}} 
\nabla_{\dot Dd} \nabla_{\dot Cc} \rho_{\underline{b}}^{\dot G}
-2 \varepsilon_{\dot B \dot A} \left(\nabla_{\dot Dd} \nabla_{\dot Cc} 
Y_{\dot E \dot F E \underline{a}} \right) \rho_{\underline{b}}^E
\nonumber \\
&-&2 \varepsilon_{\dot B \dot A} \left(\nabla_{\dot Cc} 
Y_{\dot E \dot F E \underline{a}} \right) \nabla_{\dot Dd} 
\rho_{\underline{b}}^E +2 \varepsilon_{\dot B \dot A} 
\left(\nabla_{\dot Dd} Y_{\dot E \dot F E \underline{a}} \right) 
\nabla_{\dot Cc} \rho_{\underline{b}}^E \nonumber \\
&-&2 \varepsilon_{\dot B \dot A} Y_{\dot E \dot F E \underline{a}} 
\nabla_{\dot Dd} \nabla_{\dot Cc} \rho_{\underline{b}}^E
-\varepsilon_{\dot B \dot A} \nabla_{\dot Dd} \nabla_{\dot Cc} 
\nabla^E_{\ \underline{\dot E}} \Phi_{E \underline{\dot F} ab} \nonumber \\
&+&2 \varepsilon_{\dot B \dot A} 
\Phi_{\ \underline{\dot E} \ \ \underline{a}}^{E \ \ e}
\nabla_{\dot Dd} \nabla_{\dot Cc} \Phi_{E \underline{\dot F} e \underline{b}}
-2 \varepsilon_{\dot B \dot A} \left( \nabla_{\dot Cc} 
\Phi_{\ \underline{\dot E} \ \ \underline{a}}^{E \ \ e} \right)
\nabla_{\dot Dd} \Phi_{E \underline{\dot F} e \underline{b}} \nonumber \\
&-&i \varepsilon_{ab} 
\varepsilon_{\underline{\dot E} \underline{\underline{\dot A}}}\nabla_{\dot Dd}
\nabla_{\dot Cc} \nabla^E_{\ \underline{\dot F}} 
U_{E \underline{\underline{\dot B}}} +\varepsilon_{ab} 
\varepsilon_{\underline{\dot E} \dot A} \varepsilon_{\dot B \underline{\dot F}}
\left(\nabla_{\dot Dd} U^{G \dot G} \right) \nabla_{\dot Cc} U_{G \dot G} 
\nonumber \\
&+& \varepsilon_{ab} \varepsilon_{\underline{\dot E} \dot A} 
\varepsilon_{\dot B \underline{\dot F}} U^{G \dot G} \nabla_{\dot Dd} 
\nabla_{\dot Cc} U_{G \dot G} -2i \varepsilon_{\dot B \underline{\dot E}} 
\left(\nabla_{\dot Dd} \nabla_{\dot Cc} X_{ab} \right)
W_{\underline{\dot F} \dot A} \nonumber \\
&+&2i \varepsilon_{\dot B \underline{\dot E}} \left(\nabla_{\dot Cc} X_{ab} 
\right) \nabla_{\dot Dd} W_{\underline{\dot F} \dot A} -2i 
\varepsilon_{\dot B \underline{\dot E}} \left(\nabla_{\dot Dd} X_{ab} \right)
\nabla_{\dot Cc} W_{\underline{\dot F} \dot A} \nonumber \\
&-&2i \varepsilon_{\dot B \underline{\dot E}} X_{ab} \nabla_{\dot Dd} 
\nabla_{\dot Cc} W_{\underline{\dot F} \dot A} -2i 
\varepsilon_{\dot A \underline{\dot E}} \left(\nabla_{\dot Dd} 
\nabla_{\dot Cc} X_{ab} \right) Y_{\underline{\dot F} \dot B} \nonumber \\
&+&2i \varepsilon_{\dot A \underline{\dot E}} \left(\nabla_{\dot Cc} X_{ab} 
\right) \nabla_{\dot Dd} Y_{\underline{\dot F} \dot B} -2i 
\varepsilon_{\dot A \underline{\dot E}} \left(\nabla_{\dot Dd} X_{ab} \right)
\nabla_{\dot Cc} Y_{\underline{\dot F} \dot B} \nonumber \\
&-&2i \varepsilon_{\dot A \underline{\dot E}} X_{ab} \nabla_{\dot Dd} 
\nabla_{\dot Cc} Y_{\underline{\dot F} \dot B} +\varepsilon_{ab} 
\varepsilon_{\underline{\dot E} \dot A} \varepsilon_{\dot B \underline{\dot F}}
W^{AB} \nabla_{\dot Dd} \nabla_{\dot Cc} Y_{AB} \nonumber \\
&+&2\varepsilon_{ab} \varepsilon_{\underline{\dot E} \dot A} 
\varepsilon_{\dot B \underline{\dot F}} \left(\nabla_{\dot Dd} X^{ef} \right)
\nabla_{\dot Cc} X_{ef} +2\varepsilon_{ab} 
\varepsilon_{\underline{\dot E} \dot A} \varepsilon_{\dot B \underline{\dot F}}
X^{ef} \nabla_{\dot Dd} \nabla_{\dot Cc} X_{ef} \nonumber \\
&+&\frac{1}{6} \varepsilon_{ab} \varepsilon_{\underline{\dot E} \dot A} 
\varepsilon_{\dot B \underline{\dot F}} \nabla_{\dot Dd} \nabla_{\dot Cc} R 
\label{d3dw}\\
\nabla_{\dot Dd} \nabla_{\dot Cc} \nabla_{\dot Bb} \Lambda_{\dot Aa} &=& 
-\frac{3}{2} \varepsilon_{ab} \varepsilon_{\dot B \dot A} 
\nabla_{\dot Dd} \nabla_{\dot Cc} \nabla^{F \dot F}
U_{F \dot F} -\frac{3}{2}i \varepsilon_{ab} \varepsilon_{\dot B \dot A} 
W^{AB} \nabla_{\dot Dd} \nabla_{\dot Cc} Y_{AB} \nonumber \\ 
&+&\frac{3}{2}i \varepsilon_{ab} \varepsilon_{\dot B \dot A} \left(
\nabla_{\dot Dd} W^{\dot E \dot F} \right) \nabla_{\dot Cc} Y_{\dot E \dot F} 
+\frac{3}{2}i \varepsilon_{ab} \varepsilon_{\dot B \dot A} W^{\dot E \dot F} 
\nabla_{\dot Dd} \nabla_{\dot Cc} Y_{\dot E \dot F} \nonumber \\
&+&\frac{3}{2}i \varepsilon_{ab} \varepsilon_{\dot B \dot A} \left(
\nabla_{\dot Dd} Y^{\dot E \dot F} \right) \nabla_{\dot Cc} W_{\dot E \dot F} 
+\frac{3}{2}i \varepsilon_{ab} \varepsilon_{\dot B \dot A} Y^{\dot E \dot F} 
\nabla_{\dot Dd} \nabla_{\dot Cc} W_{\dot E \dot F} \nonumber \\
&+&6 \left( \nabla_{\dot Dd} X_{ab} \right) \nabla_{\dot Cc} Y_{\dot B \dot A} 
+6 X_{ab} \nabla_{\dot Dd} \nabla_{\dot Cc} Y_{\dot B \dot A}
+6 Y_{\dot B \dot A} \nabla_{\dot Dd} \nabla_{\dot Cc} X_{ab} \nonumber \\
&+&6 \left(\nabla_{\dot Dd} Y_{\dot B \dot A} \right) \nabla_{\dot Cc} X_{ab}
+ \frac{i}{8} \varepsilon_{ab} \varepsilon_{\dot B \dot A} \nabla_{\dot Dd} 
\nabla_{\dot Cc} I
\end{eqnarray}

These results require knowing second spinorial derivatives of superfields, some
of which we have computed in section \ref{appendix31}, but others we have not 
computed yet. We present those here:
\begin{eqnarray}
\nabla_{\dot Dd} \nabla_{\dot Cc} W_{\dot E \dot F \dot A \dot B} &=&
-4i \left(\nabla_{\dot Dd} Y_{\dot B \underline{\dot A}} \right)
W_{\underline{\dot B} \underline{\dot E} \underline{\dot F} c} 
-4i Y_{\dot B \underline{\dot A}} \nabla_{\dot Dd}
W_{\underline{\dot B} \underline{\dot E} \underline{\dot F} c} \nonumber \\
&-&8 \varepsilon_{\dot C \underline{\dot F}} \left(\nabla_{\dot Dd} X_c^{\ e}
\right) W_{\underline{\dot A} \underline{\dot B} \underline{\dot E} e}
-8 \varepsilon_{\dot C \underline{\dot F}} X_c^{\ e} \nabla_{\dot Dd}
W_{\underline{\dot A} \underline{\dot B} \underline{\dot E} e} \nonumber \\
&-&2i \left(\nabla_{\dot Dd} Y_{\underline{\dot A} \underline{\dot B}}
\right) W_{\underline{\dot E} \underline{\dot F} \dot C c} -2i
Y_{\underline{\dot A} \underline{\dot B}} \nabla_{\dot Dd} 
W_{\underline{\dot E} \underline{\dot F} \dot C c} - 2i 
\varepsilon_{\dot C \underline{\dot F}} \nabla_{\dot Dd} 
\nabla^E_{\ \underline{\dot E}} Y_{\underline{\dot A} \underline{\dot B} E c} 
\nonumber \\
&+&3 \varepsilon_{\dot C \underline{\dot F}} \left(\nabla_{\dot Dd} 
U^E_{\ \underline{\dot E}}\right) Y_{\underline{\dot A} \underline{\dot B} E c}
+ 3 \varepsilon_{\dot C \underline{\dot F}} U^E_{\ \underline{\dot E}}
\nabla_{\dot Dd} Y_{\underline{\dot A} \underline{\dot B} E c} \\
\nabla_{\dot Dd} \nabla_{\dot Cc} \rho_{\dot B b}&=& -
\varepsilon_{\dot C \dot B} \nabla_{\dot Dd} X_{cb} -i \varepsilon_{cb} 
\nabla_{\dot Dd} Y_{\dot B \dot C} +2 \rho_{\dot B c} \nabla_{\dot Dd} 
\rho_{\dot C b} -2 \rho_{\dot C b} \nabla_{\dot Dd} \rho_{\dot B c} \\
\nabla_{\dot Dd} \nabla_{\dot Cc} \rho_{B b}&=&\frac{i}{4} \varepsilon_{cb}
\nabla_{\dot Dd} H_{B \dot C} - \varepsilon_{cb} \nabla_{\dot Dd} U_{B \dot C}
+i \nabla_{\dot Dd} \Phi_{B \dot C cb} +2 \rho_{B c} \nabla_{\dot Dd} 
\rho_{\dot C b} \nonumber \\
&-&2 \rho_{\dot C b} \nabla_{\dot Dd} \rho_{B c} \\
\nabla_{\dot Dd} \nabla_{\dot Cc} Y_{\dot A \dot B A a} &=& \varepsilon_{ac}
\nabla_{\dot Dd} \nabla_{A \underline{\dot A}} Y_{\dot C \underline{\dot B}}
+2i \varepsilon_{ac} \left(\nabla_{\dot Dd} U_{A \underline{\dot A}} \right) 
Y_{\dot C \underline{\dot B}} +2i \varepsilon_{ac} U_{A \underline{\dot A}} 
\nabla_{\dot Dd} Y_{\dot C \underline{\dot B}} \nonumber \\
&-& i \varepsilon_{ac} \left(\nabla_{\dot Dd} U_{A \dot C} \right) 
Y_{\dot A \dot B} -i \varepsilon_{ac} U_{A \dot C} \nabla_{\dot Dd}
Y_{\dot A \dot B} - \varepsilon_{ac} \varepsilon_{\dot C \underline{\dot B}} 
\nabla_{\dot Dd} \nabla^B_{\ \underline{\dot A}} W_{AB} \nonumber \\
&+&i \varepsilon_{\dot C \underline{\dot B}} \nabla_{\dot Dd} 
\nabla_{A \underline{\dot A}} X_{ac} -2 \varepsilon_{\dot C \underline{\dot B}}
\left(\nabla_{\dot Dd} U_{A \underline{\dot A}} \right) X_{ac} \nonumber \\
&-& 2 \varepsilon_{\dot C \underline{\dot B}} U_{A \underline{\dot A}} 
\nabla_{\dot Dd} X_{ac} \\
\nabla_{\dot Dd} \nabla_{\dot Cc} \Phi_{B \dot B ba} &=& 2i 
\varepsilon_{\underline{b}c} \varepsilon_{\dot B \dot C} \nabla_{\dot Dd}
W_{BA \ \underline{a}}^{\ \ \ A} +\frac{2}{3} \varepsilon_{\underline{b}c} 
\varepsilon_{\dot B \dot C} \nabla_{\dot Dd} \Lambda_{B \underline{a}}
+2 \varepsilon_{\underline{b}c} \varepsilon_{\dot B \dot C} \left(
\nabla_{\dot Dd} \rho^A_{\underline{a}} \right) W_{AB} \nonumber \\
&-&2i \varepsilon_{\underline{b}c} \nabla_{\dot Dd}
Y_{\dot B \dot C B\underline{a}} +2 \varepsilon_{c\underline{b}} \left(
\nabla_{\dot Dd} \rho_{B \underline{a}} \right) Y_{\dot B \dot C}
-2 \varepsilon_{c\underline{b}} \rho_{B \underline{a}} \nabla_{\dot Dd}
Y_{\dot B \dot C} \nonumber \\
&-&i \varepsilon_{c\underline{b}} \left(\nabla_{\dot Dd} 
\rho_{\dot C\underline{a}} \right) U_{B \dot B} +i \varepsilon_{c\underline{b}}
\rho_{\dot C\underline{a}} \nabla_{\dot Dd} U_{B \dot B} +2i
\varepsilon_{c\underline{b}} \left(\nabla_{\dot Dd} \rho_{\dot B\underline{a}}
\right) U_{B \dot C} \nonumber \\
&-& 2i \varepsilon_{c\underline{b}} \rho_{\dot B \underline{a}}\nabla_{\dot Dd}
U_{B \dot C} +4 \varepsilon_{c\underline{b}} \left(\nabla_{\dot Dd} 
\rho_{\dot C}^e \right) \Phi_{B \dot B e \underline{a}} -4 
\varepsilon_{c\underline{b}} \rho_{\dot C}^e \nabla_{\dot Dd} 
\Phi_{B \dot B e \underline{a}} \nonumber \\
&-&2 \varepsilon_{c\underline{b}} \nabla_{\dot Dd} \nabla_{B \dot B}
\rho_{\dot C\underline{a}} -2i \varepsilon_{\dot B \dot C} \left(
\nabla_{\dot Dd} \rho_{B \underline{b}} \right) X_{c \underline{a}} +2i 
\varepsilon_{\dot B \dot C} \rho_{B \underline{b}} \nabla_{\dot Dd} 
X_{c \underline{a}} \nonumber \\
&+&2 \left(\nabla_{\dot Dd} \rho_{\dot C\underline{b}} \right) 
\Phi_{B \dot B c \underline{a}} -2 \rho_{\dot C\underline{b}} 
\nabla_{\dot Dd} \Phi_{B \dot B c \underline{a}} \\
\nabla_{\dot Dd} \nabla_{\dot Cc} U_{B \dot B} &=& -\nabla_{\dot Dd} 
Y_{\dot B \dot C Bc} -\varepsilon_{\dot B \dot C} \nabla_{\dot Dd} 
W_{AB\ c}^{\ \ \ A} +\frac{2}{3}i \varepsilon_{\dot B \dot C}
\nabla_{\dot Dd} \Lambda_{Bc} \\
\nabla_{\dot Dd} \nabla_{\dot Cc} R &=& 4i\nabla_{\dot Dd} 
\nabla^B_{\ \dot C} W_{B A \ c}^{\ \ \ A} 
+12 \left(\nabla_{\dot Dd} X_{cb} \right) W_{\dot C \dot B}^{\ \ \ \dot B b} 
+12 X_{cb} \nabla_{\dot Dd} W_{\dot C \dot B}^{\ \ \ \dot B b} \nonumber \\
&-&2i W^{CB} \nabla_{\dot Dd} Y_{CB \dot C c}
- 6i\left(\nabla_{\dot Dd} W_{\dot A \dot B \dot C c} \right) Y^{\dot A \dot B}
+6iW_{\dot A \dot B \dot C c} \nabla_{\dot Dd} Y^{\dot A \dot B} \nonumber \\
&+& 12 \left(\nabla_{\dot Dd} U^{B \dot B} \right) Y_{\dot C \dot B B c} 
+ 12 U^{B \dot B} \nabla_{\dot Dd} Y_{\dot C \dot B B c} + 4 \left(
\nabla_{\dot Dd} U^B_{\ \dot C} \right) W_{B C \ c}^{\ \ \ C} \nonumber \\
&+& 4 U^B_{\ \dot C} \nabla_{\dot Dd} W_{B C \ c}^{\ \ \ C} \nonumber \\
\nabla_{\dot Dd} \nabla_{\dot Cc} I &=& 4\nabla_{\dot Dd} \nabla_{\dot Cc} R
-48 \left( \nabla_{\dot Dd} X^{ab} \right) \nabla_{\dot Cc} X_{ab} 
-48 X^{ab} \nabla_{\dot Dd} \nabla_{\dot Cc} X_{ab} \nonumber \\
&-&12 W^{AB} \nabla_{\dot Dd} \nabla_{\dot Cc} Y_{AB} -12 \left(
\nabla_{\dot Dd} W^{\dot A \dot B} \right) \nabla_{\dot Cc} Y_{\dot A \dot B}
\nonumber \\
&-& 12 W^{\dot A \dot B} \nabla_{\dot Dd} \nabla_{\dot Cc} Y_{\dot A \dot B}
-12 \left(\nabla_{\dot Dd} Y^{\dot A \dot B} \right) \nabla_{\dot Cc} 
W_{\dot A \dot B} \nonumber \\
&-& 12 Y^{\dot A \dot B} \nabla_{\dot Dd} \nabla_{\dot Cc} W_{\dot A \dot B} 
- 24 \left( \nabla_{\dot Dd} U^{F \dot F} \right) \nabla_{\dot Cc} 
U_{F \dot F} \nonumber \\
&-& 24 U^{F \dot F} \nabla_{\dot Dd} \nabla_{\dot Cc} U_{F \dot F} 
+ 3 P \nabla_{\dot Dd} \nabla_{\dot Cc} \overline{P}
+ 3 \left( \nabla_{\dot Dd} H^{F \dot F} \right) \nabla_{\dot Cc} 
H_{F \dot F} \nonumber \\
&+& 3 H^{F \dot F} \nabla_{\dot Dd} \nabla_{\dot Cc} H_{F \dot F}
-24 \left( \nabla_{\dot Dd} \Phi^{F \dot F}_{ab} \right) \nabla_{\dot Cc} 
\Phi^{ab}_{F \dot F} \nonumber \\
&-& 24 \Phi^{F \dot F}_{ab} \nabla_{\dot Dd} \nabla_{\dot Cc} 
\Phi^{ab}_{F \dot F} -6 \nabla_{\dot Dd} \nabla_{\dot Cc} \nabla^{F \dot F} 
H_{F \dot F} \nonumber \\
&-& 16i \left(\nabla_{\dot Dd} \nabla_{\dot Cc} \rho^{Aa} \right) \Lambda_{Aa} 
- 16i \left(\nabla_{\dot Cc} \rho^{Aa} \right) \nabla_{\dot Dd} \Lambda_{Aa}
\nonumber \\
&+&16i \left(\nabla_{\dot Dd} \rho^{Aa} \right) \nabla_{\dot Cc} \Lambda_{Aa} 
- 16i \rho^{Aa} \nabla_{\dot Dd} \nabla_{\dot Cc} \Lambda_{Aa} \nonumber \\
&+& 16i \left(\nabla_{\dot Dd} \nabla_{\dot Cc} \rho^{\dot Aa} \right) 
\Lambda_{\dot Aa}
+16i \left(\nabla_{\dot Cc} \rho^{\dot Aa} \right) \nabla_{\dot Dd}
\Lambda_{\dot Aa} \nonumber \\
&-&16i \left(\nabla_{\dot Dd} \rho^{\dot Aa} \right) \nabla_{\dot Cc} 
\Lambda_{\dot Aa} + 16i \rho^{\dot Aa} \nabla_{\dot Dd} \nabla_{\dot Cc} 
\Lambda_{\dot Aa} \nonumber \\
&+& 48 \left(\nabla_{\dot Dd} \nabla_{\dot Cc} \rho^{Aa} \right) 
W_{AB\ a}^{\ \ \ B} + 48 \left(\nabla_{\dot Cc} \rho^{Aa} \right) 
\nabla_{\dot Dd} W_{AB\ a}^{\ \ \ B} \nonumber \\
&-&48 \left(\nabla_{\dot Dd} \rho^{Aa} \right) \nabla_{\dot Cc} 
W_{AB\ a}^{\ \ \ B} +48 \rho^{Aa} \nabla_{\dot Dd} \nabla_{\dot Cc} 
W_{AB\ a}^{\ \ \ B} \nonumber \\
&-& 48\left(\nabla_{\dot Dd} \nabla_{\dot Cc} \rho^{\dot Aa} \right) 
W_{\dot A \dot B\ a}^{\ \ \ \dot B} -48 \left(\nabla_{\dot Cc} \rho^{\dot Aa} 
\right) \nabla_{\dot Dd} W_{\dot A \dot B\ a}^{\ \ \ \dot B} \nonumber \\
&+&48\left(\nabla_{\dot Dd} \rho^{\dot Aa} \right) \nabla_{\dot Cc} 
W_{\dot A \dot B\ a}^{\ \ \ \dot B} -48 \rho^{\dot Aa} \nabla_{\dot Dd} 
\nabla_{\dot Cc} W_{\dot A \dot B\ a}^{\ \ \ \dot B} \nonumber \\
&-&96i W^{AB} \left(\nabla_{\dot Dd} \rho_{Aa} \right) \nabla_{\dot Cc} 
\rho^a_B +96i W^{AB} \rho_{Aa} \nabla_{\dot Dd} \nabla_{\dot Cc} \rho^a_B 
\nonumber \\
&-&96i \left(\nabla_{\dot Dd} W^{\dot A \dot B} \right) \rho_{\dot Aa} 
\nabla_{\dot Cc} \rho^a_{\dot B} -96i W^{\dot A \dot B} \left(\nabla_{\dot Dd} 
\rho_{\dot Aa} \right) \nabla_{\dot Cc} \rho^a_{\dot B} \nonumber \\ 
&+&96i W^{\dot A \dot B} \rho_{\dot Aa} \nabla_{\dot Dd} \nabla_{\dot Cc} 
\rho^a_{\dot B} -96i \rho_{\dot Aa} \left(\nabla_{\dot Dd} 
\rho^a_{\dot B} \right) \nabla_{\dot Cc} W^{\dot A \dot B} \nonumber \\
&+&48i \rho_{\dot Aa} \rho^a_{\dot B} \nabla_{\dot Dd} \nabla_{\dot Cc} 
W^{\dot A \dot B} -48 \left(\nabla_{\dot Dd} \nabla_{\dot Cc} \rho_{Aa} 
\right) \rho^a_{\dot A} U^{A \dot A} \nonumber \\
&-&48 U^{A \dot A} \left(\nabla_{\dot Cc} \rho_{Aa} \right) \nabla_{\dot Dd} 
\rho^a_{\dot A} +48 \left(\nabla_{\dot Cc} \rho_{Aa} \right) \rho^a_{\dot A}
\nabla_{\dot Dd} U^{A \dot A} \nonumber \\
&+&48 \left(\nabla_{\dot Dd} U^{A \dot A} \right) \rho_{Aa} 
\nabla_{\dot Cc} \rho^a_{\dot A} +48 U^{A \dot A} \left(\nabla_{\dot Dd} 
\rho_{Aa} \right) \nabla_{\dot Cc} \rho^a_{\dot A} \nonumber \\
&-&48 U^{A \dot A} \rho_{Aa} \nabla_{\dot Dd} \nabla_{\dot Cc} 
\rho^a_{\dot A} -48 \left(\nabla_{\dot Dd} \rho_{Aa} \right) \rho^a_{\dot A} 
\nabla_{\dot Cc} U^{A \dot A} \nonumber \\
&+&48 \rho_{Aa} \left(\nabla_{\dot Dd} \rho^a_{\dot A} \right) \nabla_{\dot Cc}
U^{A \dot A} -48 \rho_{Aa} \rho^a_{\dot A} \nabla_{\dot Dd} \nabla_{\dot Cc} 
U^{A \dot A} \nonumber \\
&+&48i \left(\nabla_{\dot Dd} \nabla_{\dot Cc} \rho_{Aa} \right) 
\nabla^{A \dot A} \rho^a_{\dot A} +48i \left(\nabla_{\dot Cc} \rho_{Aa} \right)
\nabla_{\dot Dd} \nabla^{A \dot A} \rho^a_{\dot A} \nonumber \\
&-& 48i \left(\nabla_{\dot Dd} \rho_{Aa} \right) \nabla_{\dot Cc} 
\nabla^{A \dot A} \rho^a_{\dot A} +48i \rho_{Aa} \nabla_{\dot Dd}
\nabla_{\dot Cc} \nabla^{A \dot A} \rho^a_{\dot A} \nonumber \\
&-&48i \left(\nabla_{\dot Dd} \nabla_{\dot Cc} \rho_{\dot Aa} \right) 
\nabla^{A \dot A} \rho^a_A -48i \left(\nabla_{\dot Cc} \rho_{\dot A a} \right)
\nabla_{\dot Dd} \nabla^{A \dot A} \rho^a_A \nonumber \\
&+&48i \left(\nabla_{\dot Dd} \rho_{\dot Aa} \right) \nabla_{\dot Cc}
\nabla^{A \dot A} \rho^a_A -48i \rho_{\dot Aa} \nabla_{\dot Dd} 
\nabla_{\dot Cc} \nabla^{A \dot A} \rho^a_A \nonumber \\
&+&96i \left(\nabla_{\dot Dd} \Phi^{ab}_{A \dot A} \right) \rho^{\dot A}_b 
\nabla_{\dot Cc} \rho^A_a +96i \Phi^{ab}_{A \dot A} \left(\nabla_{\dot Dd} 
\rho^{\dot A}_b \right) \nabla_{\dot Cc} \rho^A_a \nonumber \\
&-&96i \Phi^{ab}_{A \dot A} \rho^{\dot A}_b \nabla_{\dot Dd} \nabla_{\dot Cc} 
\rho^A_a +96i \left(\nabla_{\dot Dd} \Phi^{ab}_{A \dot A} \right) \rho^A_b 
\nabla_{\dot Cc} \rho^{\dot A}_a \nonumber \\
&+&96i \Phi^{ab}_{A \dot A} \left(\nabla_{\dot Dd} \rho^A_b \right) 
\nabla_{\dot Cc} \rho^{\dot A}_a -96i \Phi^{ab}_{A \dot A} \rho^A_b 
\nabla_{\dot Dd} \nabla_{\dot Cc} \rho^{\dot A}_a \nonumber \\
&+&96i \left(\nabla_{\dot Dd} \rho^A_a \right) \rho^{\dot A}_b 
\nabla_{\dot Cc} \Phi^{ab}_{A \dot A} - 96i \rho^A_a \left(\nabla_{\dot Dd} 
\rho^{\dot A}_b \right) \nabla_{\dot Cc} \Phi^{ab}_{A \dot A} \nonumber \\
&+&96i \rho^A_a \rho^{\dot A}_b \nabla_{\dot Dd} \nabla_{\dot Cc} 
\Phi^{ab}_{A \dot A} \label{ddi}
\end{eqnarray}

Because of the $\nabla_{\dot Dd} \nabla_{\dot Cc} I$ term in 
$\nabla_{\dot Dd} \nabla_{\dot Cc} \nabla_{\dot Bb} \Lambda_{\dot Aa}$, we will
also need the following terms:
\begin{eqnarray}
\nabla_{\dot Dd} \nabla_{\dot Cc} \overline{P} &=&
-8i \nabla_{\dot Dd} W^{\ \ \ \dot B}_{\dot C \dot B \ c} 
-\frac{4}{3} \nabla_{\dot Dd} \Lambda_{\dot C c} +8 \left(\nabla_{\dot Dd}
\rho^{\dot B}_c \right) W_{\dot C \dot B} 
-8 \rho^{\dot B}_c \nabla_{\dot Dd} W_{\dot C \dot B} \nonumber \\
&+&2 \left(\nabla_{\dot Dd} \overline{P} \right) \rho_{\dot C c}
+2 \overline{P} \nabla_{\dot Dd} \rho_{\dot C c} +2 \left(\nabla_{\dot Dd} 
\rho^C_c \right) H_{C \dot C} 
-2 \rho^C_c \nabla_{\dot Dd} H_{C \dot C} \nonumber \\
&+&4i \left(\nabla_{\dot Dd} \rho^C_c \right) U_{C \dot C} -4i \rho^C_c 
\nabla_{\dot Dd} U_{C \dot C} -8 \nabla_{\dot Dd} \nabla_{C \dot C} \rho^C_c 
\nonumber \\
&+& 8 \left(\nabla_{\dot Dd} \Phi_{C\dot C cb} \right) \rho^{Cb}
+8 \Phi_{C\dot C cb} \nabla_{\dot Dd} \rho^{Cb}\\
\nabla_{\dot Dd} \nabla_{\dot Cc} H_{B \dot B} &=&8i 
\varepsilon_{\dot C \dot B} \nabla_{\dot Dd} W_{AB\ c}^{\ \ \ A} +\frac{4}{3}
\varepsilon_{\dot C \dot B} \nabla_{\dot Dd} \Lambda_{B c}
+4 \varepsilon_{\dot C \dot B} W_{BC} \nabla_{\dot Dd} \rho^C_c \nonumber \\
&-&4i \varepsilon_{\dot C \dot B} X_{bc} \nabla_{\dot Dd} \rho^b_B
+4i \varepsilon_{\dot C \dot B} \rho^b_B \nabla_{\dot Dd} X_{bc} -2 
\varepsilon_{\dot C \dot B} P \nabla_{\dot Dd} \rho_{B c} \nonumber \\
&-&2 \varepsilon_{\dot C \dot B} H_{B \dot A} \nabla_{\dot Dd} \rho_c^{\dot A} 
+2 \varepsilon_{\dot C \dot B} \rho_c^{\dot A} \nabla_{\dot Dd} H_{B \dot A}
-4 Y_{\dot C \dot B} \nabla_{\dot Dd} \rho_{B c} \nonumber \\
&+&4 \rho_{B c} \nabla_{\dot Dd} Y_{\dot C \dot B} -2i U_{B \dot B} 
\nabla_{\dot Dd} \rho_{\dot Cc}+2i \rho_{\dot Cc} \nabla_{\dot Dd} U_{B \dot B}
\nonumber \\
&+&8 \Phi_{B \dot B bc} \nabla_{\dot Dd} \rho^b_{\dot C} -8 \rho^b_{\dot C} 
\nabla_{\dot Dd} \Phi_{B \dot B bc} -4 \nabla_{\dot Dd} \nabla_{B \dot B} 
\rho_{\dot Cc} \nonumber \\
&+&8 \nabla_{\dot Dd} \nabla_{B \dot C} \rho_{\dot Bc}
+8 \rho^b_{\dot B} \nabla_{\dot Dd} \Phi_{B \dot C bc}-8 \Phi_{B \dot C bc} 
\nabla_{\dot Dd} \rho^b_{\dot B} \\
\nabla_{\dot Dd} \nabla_{\dot Cc} \Lambda_{B b} &=& 3i \varepsilon_{cb}
\nabla_{\dot Dd} \nabla_{B \dot B} W_{\dot C}^{\ \dot B} +3i \varepsilon_{cb}
\nabla_{\dot Dd} \nabla^A_{\ \dot C} Y_{AB} -3 \nabla_{\dot Dd} 
\nabla_{B \dot C} X_{cb} \nonumber \\
&+& 6i \left(\nabla_{\dot Dd} U_{B \dot C}\right) X_{cb} +6i U_{B \dot C} 
\nabla_{\dot Dd} X_{cb} \\
\nabla_{\dot Dd} \nabla_{\dot Cc} W_{ABCc} &=& - \varepsilon_{cb} 
\nabla_{\dot Dd} \nabla_{\underline{A} \dot C} W_{\underline{B}C} +2i 
\varepsilon_{cb} \left(\nabla_{\dot Dd} U_{\underline{A} \dot C}\right) 
W_{\underline{B}C} -\varepsilon_{cb} \varepsilon_{\underline{B}C} 
\nabla_{\dot Dd} \nabla_{\underline{A}}^{\ \dot B} Y_{\dot B \dot C} 
\nonumber \\
&-&i \varepsilon_{\underline{B}C} \nabla_{\dot Dd} 
\nabla_{\underline{A} \dot C} X_{cb} -i \varepsilon_{cb} \left(
\nabla_{\dot Dd} U_{C \dot C} \right) W_{AB} -2 \varepsilon_{\underline{B}C}
\left(\nabla_{\dot Dd} U_{\underline{A} \dot C} \right) X_{cb} \nonumber \\
&-&2 \varepsilon_{\underline{B}C} U_{\underline{A} \dot C} \nabla_{\dot Dd} 
X_{cb} \\
\nabla_{\dot Dd} \nabla_{\dot Cc} Y_{AB} &=& 2i \nabla_{\dot Dd} Y_{AB \dot Cc}
\\
\nabla_{\dot Dd} \nabla_{\dot Cc} Y_{\dot A \dot B} &=& \frac{2}{3}
\varepsilon_{\dot C \underline{\dot B}} \nabla_{\dot Dd} 
\Lambda_{\underline{\dot A} c} \\
\nabla_{\dot Dd} \nabla_{\dot Cc} X_{ab} &=& -\frac{2}{3}i \varepsilon_{c
\underline{a}} \nabla_{\dot Dd} \Lambda_{\dot C \underline{b}}
\end{eqnarray}
All other differential relations we will need involve combinations of vector
and spinor covariant derivatives. Terms containing these expressions are very 
important for our analysis, since a lot of the derivatives we are looking for 
come from them. They may be written, using the commutation relations, as 
vector derivatives of the relations we have seen plus torsion and curvature 
terms. Some of the torsion terms require differential relations involving 
undotted and dotted spinorial derivatives which we compute now:
\begin{eqnarray}
\nabla_A^a \nabla_{\dot C}^c \rho_{\dot B}^b&=& -\frac{i}{4} 
\varepsilon_{\dot C \dot B} \varepsilon^{cb} \nabla_A^a P -
\varepsilon_{\dot C \dot B} \nabla_A^a X^{cb} -i \varepsilon^{cb} \nabla_A^a
Y_{\dot C \dot B} +2 \rho_{\dot B}^c \nabla_A^a \rho_{\dot C}^b \nonumber \\
&-&2 \rho_{\dot C}^b \nabla_A^a \rho_{\dot B}^c \\
\nabla_A^a \nabla_{\dot C}^c \rho_B^b &=&\frac{i}{4} \varepsilon^{cb}
\nabla_A^a H_{B \dot C} -\varepsilon^{cb} \nabla_A^a U_{B \dot C}
+i \nabla_A^a \Phi_{B \dot C}^{bc} +2 \rho_B^c \nabla_A^a \rho_{\dot C}^b 
\nonumber \\
&-&2 \rho_{\dot C}^b \nabla_A^a \rho_B^c \\
\nabla_D^d \nabla_{\dot C}^c H_{B \dot B} &=& 8i \varepsilon_{\dot C \dot B}
\nabla_D^d W_{BC}^{\ \ \ C c} +\frac{4}{3} \varepsilon_{\dot C \dot B}
\nabla_D^d \Lambda_B^c +4 \varepsilon_{\dot C \dot B} W_{BC} \nabla_D^d 
\rho^{C c} \nonumber \\
&-&4 \varepsilon_{\dot C \dot B} \rho^{C c} \nabla_D^d W_{BC}
+4i \varepsilon_{\dot C \dot B} X^{cb} \nabla_D^d \rho^C_b 
-4i \varepsilon_{\dot C \dot B} \rho^C_b \nabla_D^d X^{cb} \nonumber \\
&-&2 \varepsilon_{\dot C \dot B} \rho^c_B \nabla_D^d P 
-2 \varepsilon_{\dot C \dot B} P \nabla_D^d \rho^c_B -2 
\varepsilon_{\dot C \dot B} H_{B \dot A} \nabla_D^d \rho^{\dot A c}\nonumber \\
&+&2 \varepsilon_{\dot C \dot B} \rho^{\dot A c} \nabla_D^d H_{B \dot A} 
-4 Y_{\dot C \dot B} \nabla_D^d \rho^c_B +4 \rho^c_B \nabla_D^d 
Y_{\dot C \dot B} -2i U_{B \dot B} \nabla_D^d \rho^c_{\dot C} \nonumber \\
&+&2i \rho^c_{\dot C} \nabla_D^d U_{B \dot B} 
-8 \Phi_{B \dot B}^{bc} \nabla_D^d \rho_{\dot C b}
+8 \rho_{\dot C b} \nabla_D^d \Phi_{B \dot B}^{bc}
+8 \Phi_{B \dot C}^{bc} \nabla_D^d \rho_{\dot B b} \nonumber \\
&-&8 \rho_{\dot B b} \nabla_D^d \Phi_{B \dot C}^{bc}
-4 \nabla_D^d \nabla_{B \dot B} \rho^c_{\dot C}
+8 \nabla_D^d \nabla_{B \dot C} \rho^c_{\dot B} \\
\nabla_{Dd} \nabla_{\dot Cc} U_{B \dot B} &=& -\nabla_{Dd} Y_{\dot C \dot B Bc}
+\varepsilon_{\dot C \dot B} \nabla_{Dd} W_{BC \ c}^{\ \ \ C} -\frac{2}{3}i
\varepsilon_{\dot C \dot B} \nabla_{Dd} \Lambda_{Bc}  \\
\nabla_D^d \nabla_{\dot C}^c \Phi_{B \dot B}^{ab} &=& 2i 
\varepsilon_{\dot C \dot B} \varepsilon^{c\underline{b}} \nabla_D^d 
W_{BC}^{\ \ \ C \underline{a}} +\frac{2}{3} \varepsilon_{\dot C \dot B} 
\varepsilon^{c\underline{b}} \nabla_D^d \Lambda_B^{\underline{a}}
+2 \varepsilon_{\dot C \dot B} \varepsilon^{c\underline{b}} W_{BC}
\nabla_D^d \rho^{C \underline{a}} \nonumber \\
&-&2 \varepsilon_{\dot C \dot B} \varepsilon^{c\underline{b}} 
\rho^{C \underline{a}} \nabla_D^d W_{BC} +2i \varepsilon^{c\underline{b}} 
\nabla_D^d Y_{\dot C \dot B B}^{\ \ \ \ \ \underline{a}} 
+2 \varepsilon^{c\underline{b}} Y_{\dot C \dot B} \nabla_D^d 
\rho_B^{\underline{a}} \nonumber \\
&-&2 \varepsilon^{c\underline{b}} \rho_B^{\underline{a}} \nabla_D^d 
Y_{\dot C \dot B} -i \varepsilon^{c\underline{b}} U_{B \dot B} \nabla_D^d 
\rho_{\dot C}^{\underline{a}} +i \varepsilon^{c\underline{b}} 
\rho_{\dot C}^{\underline{a}} \nabla_D^d U_{B \dot B} +2i 
\varepsilon^{c\underline{b}} U_{B \dot C} \nabla_D^d 
\rho_{\dot B}^{\underline{a}} \nonumber \\
&-&2i \varepsilon^{c\underline{b}} \rho_{\dot B}^{\underline{a}} 
\nabla_D^d U_{B \dot C} +2i \varepsilon_{\dot C \dot B} X^{c\underline{b}} 
\nabla_D^d \rho_B^{\underline{a}} -2i \varepsilon_{\dot C \dot B} 
\rho_B^{\underline{a}} \nabla_D^d X^{c\underline{b}} +2 
\Phi_{B \dot B}^{\underline{b}c} \nabla_D^d \rho_{\dot C}^{\underline{a}} 
\nonumber \\
&-&2 \rho_{\dot C}^{\underline{a}} \nabla_D^d \Phi_{B \dot B}^{\underline{b}c}
-4 \varepsilon^{c\underline{b}} \Phi_{B \dot B}^{\underline{a}e} \nabla_D^d 
\rho_{\dot C e} +4 \varepsilon^{c\underline{b}} \rho_{\dot C e} \nabla_D^d 
\Phi_{B \dot B}^{\underline{a}e} -2 \varepsilon^{c\underline{b}} 
\nabla_D^d \nabla_{B \dot B} \rho_{\dot C}^{\underline{a}} \nonumber \\
\end{eqnarray}

All these expressions should be enough for the direct computation of 
(\ref{d4w2}).

\section{Calculation of the derivative terms}
\label{appendix4}
\indent

In this appendix, we compute the terms arising in the calculation of
$\nabla_{\dot Cc} \nabla_{\dot Bb} \nabla_{\dot Aa} \overline{W}^2$ and
$\nabla_{\dot Dd} \nabla_{\dot Cc} \nabla_{\dot Bb} \nabla_{\dot Aa} 
\overline{W}^2$ which contain vector derivatives of the auxiliary fields 
$P, H_m$ (excluding $\nabla^m H_m$, as justified on the text). For each 
expression, we indicate its content in terms of the derivatives of interest.
The derivatives of some of these expressions will also be necessary; for 
those expressions, we give their field content in terms of $P, H_m$.
Through all this appendix, we will be only interested in the derivatives we  
mentioned and, therefore, we will only consider here expressions which contain 
them. Those expressions we do not consider here simply do not contain such 
derivatives, as it can be verified using their expansions in appendix 
\ref{appendix3}.

\subsection{Calculation of the derivative terms in $\nabla_{\dot Cc} 
\nabla_{\dot Bb} \nabla_{\dot Aa} \overline{W}^2$}
\label{appendix41}
\indent

From (\ref{d3w2}), the following expressions are necessary:
\begin{eqnarray}
\nabla_{\dot C}^c \nabla_{\dot B}^b W_{\dot D \dot E} &=& \frac{i}{4} 
\varepsilon^{bc} \varepsilon_{\underline{\dot D} \dot B} 
\varepsilon_{\underline{\dot E} \dot C} \left(P \overline{P}+H^2-
2 \nabla_{A \dot A} H^{A \dot A} \right) +\cdots\\
\nabla_{\dot C}^c \nabla_{\dot B}^b W_{\dot D \dot E \dot A}^{\ \ \ \ a} &=& 
-\frac{i}{2} \varepsilon_{\dot B \dot A} \varepsilon^{\underline{a}c} P
W_{\dot D \dot E \dot C}^{\ \ \ \ \underline{b}} +\frac{i}{2} 
\varepsilon_{\dot B \dot A} \varepsilon^{\underline{a}c} H_{E \dot C} 
Y_{\dot D \dot E}^{\ \ \ E \underline{b}} +\cdots\\
\nabla_{\dot B}^b \nabla_{A \dot A} \rho^c_{\dot C}&=&-\frac{i}{4}
\varepsilon^{bc} \varepsilon_{\dot B \dot C} \nabla_{A \dot A} P
+\frac{i}{4} \varepsilon^{bc} Y_{\dot A \dot B} H_{A \dot C} +\frac{i}{4} 
\varepsilon^{bc} \varepsilon_{\dot A \dot B} W_A^{\ D}H_{D \dot C} \nonumber \\
&-&\frac{1}{4} \varepsilon_{\dot A \dot C} \varepsilon^{bc} P U_{A \dot B} 
+\frac{1}{8} \varepsilon^{bc} \varepsilon_{\dot B \dot C} P U_{A \dot A}
- \frac{1}{4} \varepsilon_{\dot A \dot B} X^{bc} H_{A \dot C} +\cdots\\
\nabla_{\dot C}^c \nabla_{A \dot A} \rho^{Aa}&=& \frac{i}{4} \varepsilon^{ca}
\nabla_{A \dot A} H^A_{\ \dot C} -\frac{i}{2} \varepsilon^{ca} \overline{P}
Y_{\dot A \dot C} +\frac{1}{2} \varepsilon_{\dot A \dot C} \overline{P}
X^{ca} \nonumber \\
&-& \frac{1}{8} \varepsilon^{ca} U_{A \dot A} H^A_{\ \dot C}
+ \frac{1}{4} \varepsilon^{ca} U_{A \dot C} H^A_{\ \dot A} +\cdots\\
\nabla_{A \dot A} \nabla_{\dot B}^b H_{C \dot C}&=& 2 
\varepsilon_{\dot C \dot B} \rho_C^b \nabla_{A \dot A} P +2 
\varepsilon_{\dot C \dot B} P \nabla_{A \dot A} \rho_C^b +2 
\varepsilon_{\dot C \dot B} \rho^{\dot Db} \nabla_{A \dot A} H_{C \dot D} 
\nonumber \\
&+&2\varepsilon_{\dot C \dot B} H_{C \dot D} \nabla_{A \dot A} \rho^{\dot Db}
+\cdots
\end{eqnarray}

\subsection{Calculation of the derivative terms in $\nabla_{\dot Dd} 
\nabla_{\dot Cc} \nabla_{\dot Bb} \nabla_{\dot Aa} \overline{W}^2$}
\label{appendix42}
\indent

In order to compute the derivatives in (\ref{d4w2}), besides the previous 
expressions we also need
\begin{eqnarray}
\nabla_{\dot D}^d \nabla_{\dot C}^c \overline{P} &=& -2i \varepsilon^{dc}
\nabla_{C \underline{\dot C}} H^C_{\ \underline{\dot D}} +\cdots\\
\nabla_{\dot D}^d \nabla_{\dot C}^c H_{B \dot B} &=& 2i \varepsilon^{dc}
\varepsilon_{\dot B \underline{\dot C}} \nabla_{B \underline{\dot D}} P 
-2 \varepsilon_{\dot B \underline{\dot C}} X^{cd} H_{B \underline{\dot D}}
-i \varepsilon_{\dot D \dot C} \varepsilon^{dc} W_B^{\ C} H_{C \dot B} 
\nonumber \\
&-&i \varepsilon_{\dot D \dot C} \varepsilon^{dc} Y_{\dot B}^{\ \dot A} 
H_{B \dot A} +4i \varepsilon^{dc} H_{B \dot B} Y_{\dot C \dot D}
+\cdots \\
\nabla_D^d \nabla_{\dot C}^c H_{B \dot B} &=& -i \varepsilon^{dc}
\varepsilon_{\dot C \dot B} \varepsilon_{DB} \nabla_{A \dot A} H^{A \dot A}
+i \varepsilon^{dc} \nabla_{B \dot B} H_{D \dot C} -2i \varepsilon^{dc} 
\nabla_{B \dot C} H_{D \dot B}\nonumber \\
&+&\cdots \\
\nabla_{\dot D}^d \nabla_{A \dot A} \nabla_{\dot C}^c H_{B \dot B} &=&
i \varepsilon_{\dot C \dot B} \varepsilon_{AB} \varepsilon^{dc} 
Y_{\dot A \dot D} \nabla_{E \dot E} H^{E \dot E} -i \varepsilon^{dc}
Y_{\dot A \dot D} \nabla_{B \dot B} H_{A \dot C} \nonumber \\
&+&2i \varepsilon^{dc} Y_{\dot A \dot D} \nabla_{B \dot C} H_{A \dot B} 
+i \varepsilon_{\dot C \dot B} \varepsilon_{\dot A \dot D} \varepsilon^{dc} 
W_{AB} \nabla_{E \dot E} H^{E \dot E} \nonumber \\
&-&i \varepsilon_{\dot A \dot D} \varepsilon^{dc} W_A^{\ C} \nabla_{B \dot B} 
H_{C \dot C} +2i \varepsilon_{\dot A \dot D} \varepsilon^{dc} W_A^{\ C} 
\nabla_{B \dot C} H_{C \dot B} \nonumber \\
&-&\varepsilon_{\dot C \dot B} \varepsilon_{\dot A \dot D} \varepsilon_{AB} 
X^{dc} \nabla_{E \dot E} H^{E \dot E} +\varepsilon_{\dot A \dot D} X^{dc} 
\nabla_{B \dot B} H_{A \dot C} \nonumber \\
&-&2 \varepsilon_{\dot A \dot D} X^{dc} \nabla_{B \dot C} H_{A \dot B} 
-\varepsilon_{\dot B \underline{\dot C}} \varepsilon^{dc}
U_{A \dot A} \nabla_{B \underline{\dot D}} P -\varepsilon_{\dot C \dot A}
\varepsilon^{dc} U_{A \dot D} \nabla_{B \dot B} P \nonumber \\
&+&2 \varepsilon_{\dot B \dot A} \varepsilon^{dc} U_{A \dot D} 
\nabla_{B \dot C} P +2i \varepsilon_{\dot B \underline{\dot C}} 
\varepsilon^{dc} \nabla_{A \dot A} \nabla_{B \underline{\dot D}} P \nonumber \\
&-&i \varepsilon_{\dot D \dot C} \varepsilon^{dc} W_B^{\ C} \nabla_{A \dot A}
H_{C \dot B} -2 \varepsilon_{\dot B \underline{\dot C}} X^{dc} 
\nabla_{A \dot A} H_{B \underline{\dot D}} 
\nonumber \\
&-&i \varepsilon_{\dot D \dot C} 
\varepsilon^{dc} Y_{\dot B}^{\ \dot E} \nabla_{A \dot A} H_{B \dot E}
+4i \varepsilon^{dc} Y_{\dot C \dot D} \nabla_{A \dot A} H_{B \dot B}
+\cdots \\
\nabla_{\dot D}^d \nabla_{\dot C}^c \nabla_{A \dot A} H_{B \dot B} &=&
\nabla_{\dot D}^d \nabla_{A \dot A} \nabla_{\dot C}^c H_{B \dot B} 
-i \varepsilon_{\dot D \dot B} \varepsilon_{AB} \varepsilon^{dc} 
Y_{\dot A \dot C} \nabla_{E \dot E} H^{E \dot E} +i \varepsilon^{dc}
Y_{\dot A \dot C} \nabla_{B \dot B} H_{A \dot D} \nonumber \\
&-&2i \varepsilon^{dc} Y_{\dot A \dot C} \nabla_{A \dot B} H_{B \dot D} 
-i \varepsilon^{dc} \varepsilon_{\dot D \dot B} \varepsilon_{\dot A \dot C} 
W_{AB} \nabla_{E \dot E} H^{E \dot E} \nonumber \\
&+&i \varepsilon^{dc} \varepsilon_{\dot A \dot C} W_A^{\ C} \nabla_{B \dot B} 
H_{C \dot D} -2i \varepsilon^{dc} \varepsilon_{\dot A \dot C} W_A^{\ C} 
\nabla_{C \dot B} H_{B \dot D} \nonumber \\
&-& \varepsilon_{AB} \varepsilon_{\dot D \dot B} 
\varepsilon_{\dot A \dot C} X^{dc} \nabla_{E \dot E} H^{E \dot E} \nonumber \\
&+&\varepsilon_{\dot A \dot C} X^{dc} \nabla_{B \dot B} H_{A \dot D}
-2\varepsilon_{\dot A \dot C} X^{dc} \nabla_{A \dot B} H_{B \dot D}\nonumber \\
&-&\varepsilon^{dc} \varepsilon_{\dot B \underline{\dot C}} 
U_{A \dot A} \nabla_{B \underline{\dot D}} P 
+2 \varepsilon^{dc} 
\varepsilon_{\dot B \underline{\dot D}} U_{A \dot C} 
\nabla_{B \underline{\dot A}} P +\cdots
\end{eqnarray}
Remarkably, when we contract the $A, B$ and $\dot A, \dot B$ indices, lots of 
terms in the previous equations cancel by themselves. Ignoring the 
$\nabla_{E \dot E} H^{E \dot E}$ parts, in which we are not 
interested, we are simply left with
\begin{equation}
\nabla_{\dot D}^d \nabla_{\dot C}^c \nabla_{A \dot A} H^{A \dot A} =
4 \varepsilon^{dc} U^A_{\ \underline{\dot C}} 
\nabla_{A \underline{\dot D}} P +\cdots
\end{equation}
We also need more derivatives of $\rho_{\dot B}^b$:
\begin{eqnarray}
\nabla_A^a \nabla_{\dot C}^c \rho_{\dot B}^b&=& -\frac{i}{2} \varepsilon^{cb} 
\varepsilon_{\dot C \dot B} \rho_A^a P -\frac{i}{2} \varepsilon^{cb} 
H_{A \underline{\dot B}} \rho_{\underline{\dot C} a} -\frac{i}{2}
\varepsilon_{\dot C \dot B} \varepsilon^{a\underline{b}} H_{A \dot A}
\rho^{\dot A \underline{c}} \nonumber \\
&-&\frac{i}{2} \varepsilon_{\dot C \dot B}
\varepsilon^{cb} H_{A \dot A} \rho^{\dot A a} +\cdots \\
\nabla_{\dot D}^d \nabla_{\dot C}^c \nabla_{A \dot A} \rho_{\dot B}^b &=&
-\frac{i}{2} \varepsilon^{db} \varepsilon_{\dot D \dot C} \rho_{\dot B}^c 
\nabla_{A \dot A} P + \frac{i}{2} \varepsilon^{dc} \varepsilon_{\dot D \dot B} 
\rho_{\dot C}^b \nabla_{A \dot A} P +\cdots\\
\nabla_{\dot D}^d \nabla_{\dot C}^c \nabla_{A \dot A} \rho_B^b &=&
-i \varepsilon^{\underline{c}b} \varepsilon_{\dot D \dot C} 
\rho_B^{\underline{c}} \nabla_{A \dot A} P -\frac{i}{2} \varepsilon^{cb} 
\varepsilon_{\dot D \dot C} \rho^{\dot B d} \nabla_{A \dot A} H_{B \dot B}
\nonumber \\
&-&\frac{i}{2} \varepsilon^{dc} \rho_{\dot C}^b \nabla_{A \dot A} H_{B \dot D}
+\cdots
\end{eqnarray}
With respect to derivatives of $\Phi_{A \dot B}^{ab}$, we need:
\begin{eqnarray}
\nabla_{\dot D}^d \nabla_{\dot C}^c \Phi_{B \dot B}^{ab}&=& \frac{i}{2}
\varepsilon_{\dot C \dot B} \varepsilon^{c\underline{a}} 
\varepsilon^{d \underline{b}} W_B^{\ A} H_{A \dot D} +\frac{i}{2}
\varepsilon_{\dot D \dot B} \varepsilon^{c\underline{a}} 
\varepsilon^{d \underline{b}} W_B^{\ A} H_{A \dot C} +\frac{i}{2}
\varepsilon^{c\underline{a}} \varepsilon^{d \underline{b}} Y_{\dot B \dot C}
H_{B \dot D} \nonumber \\
&-&\frac{i}{2} \varepsilon^{c\underline{a}} 
\varepsilon^{d \underline{b}} Y_{\dot B \dot D} H_{B \dot C}
-\frac{1}{2} \varepsilon_{\dot C \dot B}
\varepsilon^{d\underline{b}} X^{c\underline{a}} H_{B \dot D}
-\frac{1}{2} \varepsilon_{\dot D \dot B} \varepsilon^{c\underline{b}} 
X^{d\underline{a}} H_{B \dot C} \nonumber \\
&-& i \varepsilon^{c\underline{a}}
\varepsilon_{\dot D \dot C} \Phi_{B \dot B}^{d \underline{b}} P
-\frac{i}{2} \varepsilon^{d\underline{a}} \varepsilon_{\dot D \dot C} 
\Phi_{B \dot B}^{c \underline{b}} P +\frac{i}{2} \varepsilon^{c\underline{a}} 
\varepsilon^{d\underline{b}} \varepsilon_{\dot D \dot C} \nabla_{B \dot B} P
+\cdots \label{ddphi} \\
\nabla_D^d \nabla_{\dot C}^c \Phi_{B \dot B}^{ab}&=& \frac{i}{2}
\varepsilon^{c\underline{a}} \varepsilon^{d \underline{b}} \nabla_{B \dot B}
H_{D \dot C}+\cdots\\
\nabla_{\dot D}^d \nabla_{\dot C}^c \nabla_{A \dot A} \Phi_{B \dot B}^{ab}&=& 
\frac{i}{2} \varepsilon^{c\underline{a}} \varepsilon^{d\underline{b}} 
\varepsilon_{\dot D \dot C} \nabla_{A \dot A} \nabla_{B \dot B} P \nonumber \\
&-&i \varepsilon^{c\underline{a}} \varepsilon_{\dot D \dot C} 
\Phi_{B \dot B}^{d \underline{b}} \nabla_{A \dot A} P 
-\frac{i}{2} \varepsilon^{d\underline{a}} \varepsilon_{\dot D \dot C} 
\Phi_{B \dot B}^{c \underline{b}} \nabla_{A \dot A} P \nonumber \\
&-&\frac{i}{2} \varepsilon_{\dot C \dot B} \varepsilon^{c\underline{a}} 
\varepsilon^{d \underline{b}} W_B^{\ D} \nabla_{A \dot A} H_{D \dot D}
+\frac{i}{2} \varepsilon_{\dot D \dot B} \varepsilon^{c\underline{a}} 
\varepsilon^{d \underline{b}} W_B^{\ D} \nabla_{A \dot A} H_{D \dot C}
\nonumber \\
&-&\frac{i}{2} \varepsilon_{\dot C \dot A} \varepsilon^{c\underline{a}} 
\varepsilon^{d \underline{b}} W_A^{\ D} \nabla_{B \dot B} H_{D \dot D}
+\frac{i}{2} \varepsilon_{\dot D \dot A} \varepsilon^{c\underline{a}} 
\varepsilon^{d \underline{b}} W_A^{\ D} \nabla_{B \dot B} H_{D \dot C}
\nonumber \\
&+&\frac{i}{2} \varepsilon^{c\underline{a}} \varepsilon^{d \underline{b}} 
Y_{\dot B \dot C} \nabla_{A \dot A} H_{B \dot D} -\frac{i}{2} 
\varepsilon^{c\underline{a}} \varepsilon^{d \underline{b}} Y_{\dot B \dot D} 
\nabla_{A \dot A} H_{B \dot C} \nonumber \\
&-&\frac{i}{2} \varepsilon^{c\underline{a}} \varepsilon^{d \underline{b}} 
Y_{\dot A \dot C} \nabla_{B \dot B} H_{A \dot D} +\frac{i}{2} 
\varepsilon^{c\underline{a}} \varepsilon^{d \underline{b}} Y_{\dot A \dot D} 
\nabla_{B \dot B} H_{A \dot C} \nonumber \\
&-&\frac{1}{2} \varepsilon_{\dot C \dot B} \varepsilon^{d \underline{b}} 
X^{c\underline{a}} \nabla_{A \dot A} H_{B \dot D}
-\frac{1}{2} \varepsilon_{\dot D \dot B} \varepsilon^{c \underline{b}} 
X^{d\underline{a}} \nabla_{A \dot A} H_{B \dot C} \nonumber \\
&-&\frac{1}{2} \varepsilon_{\dot C \dot A} \varepsilon^{d \underline{b}} 
X^{c\underline{a}} \nabla_{B \dot B} H_{A \dot D} 
+\frac{1}{2} \varepsilon_{\dot D \dot A} \varepsilon^{c \underline{b}} 
X^{d\underline{a}} \nabla_{B \dot B} H_{A \dot C} +\cdots \label{dddphi}
\end{eqnarray}
We did not include in (\ref{ddphi}) and (\ref{dddphi}) those terms with a 
$U_{A \dot A} \nabla_{B \dot B} P$ factor, simply because these terms exist in 
partial calculations, but overall they cancel.
The term with $\nabla_{A \dot A} \nabla_{B \dot B} P$ does not cancel in 
(\ref{dddphi}), but this expression appears in (\ref{d3dw}) as 
$\nabla_{\dot D}^d \nabla_{\dot C}^c \nabla^B_{\ \underline{\dot A}}
\Phi_{B \underline{\dot B}}^{ab}$. Therefore, this term appears as 
$\nabla^B_{\ \underline{\dot A}} \nabla_{B \underline{\dot B}} P$, a 
commutator that does not have vector derivatives.

Replacing all the appropriate expressions in (\ref{ddi}), with suitable 
contraction or symmetrization of the adequate indices, and adding all terms, 
we are led to the surprising and exciting result that $\nabla_{\dot Dd} 
\nabla_{\dot Cc} I$, like $\nabla_{\dot Cc} I$, also has no derivatives of 
$P, H_{A \dot A}$ other than $\nabla_{A \dot A} H^{A \dot A}$. We analyze 
the other terms from (\ref{d4w2}) in the main text.

\end{document}